\PassOptionsToPackage{table}{xcolor}
\PassOptionsToPackage{british}{babel}
\PassOptionsToPackage{expansion,protrusion}{microtype}
\documentclass[10pt,journal,compsoc]{IEEEtran} 
\usepackage[T1]{fontenc}
\usepackage{colortbl}
\usepackage{libertine} 
\usepackage{booktabs}
\usepackage{amsmath}
\usepackage{amssymb}
\usepackage{pxfonts}
\usepackage{hyperref}
\usepackage{url}
\usepackage{bbm}
\usepackage{cite}
\usepackage{siunitx}
\usepackage{xspace}
\usepackage{varioref}
\usepackage{flushend}
\usepackage{fontawesome}
\usepackage{makecell}
\usepackage{hyphenat}
\usepackage{mathtools} 
\usepackage[shortcuts]{extdash}
\usepackage{paralist}
\usepackage{balance}
\usepackage{tikz}
\usetikzlibrary{shapes,arrows.meta,calc,intersections,backgrounds,positioning}
\newcommand{\envelope}[3]{
  \def\ewidth{#2}
  \coordinate (#3/bl) at (#1); 

    \def\iwidth{\ewidth/3}  
    \def\eheight{0.55*\ewidth} 
    \def\lheight{0.325*\eheight} 
    \coordinate (#3/br) at ($(#3/bl) + (\ewidth,0)$); 
    \coordinate (#3/tl) at ($(#3/bl) + (90:\eheight)$); 
    \coordinate (#3/cl) at ($(#3/bl)!0.5!(#3/tl)$); 
    \coordinate (#3/tr) at ($(#3/br) + (90:\eheight)$); 
    \coordinate (#3/cr) at ($(#3/br)!0.5!(#3/tr)$); 
    \coordinate (#3/ct) at ($(#3/tl)!0.5!(#3/tr)$);  

    \coordinate (#3/fl) at ($(#3/cl) + (0:\iwidth)$); 
    \coordinate (#3/fr) at ($(#3/cr) + (0:-\iwidth)$); 

    \coordinate (#3/eil) at ($(#3/fl)!0.7!(#3/tl)$); 
    \coordinate (#3/eir) at ($(#3/fr)!0.7!(#3/tr)$); 

    \coordinate (#3/top) at ($(#3/ct) + (90:\iwidth)$);
    
    \path[envelope](#3/tl) [rounded corners=\ewidth/12] -- (#3/top)
    [sharp corners] -- (#3/tr) -- (#3/br) -| (#3/tl);

    \path[envelope] (#3/fl) -- (#3/bl) -- (#3/tl) -- cycle -- (#3/fr)
    -- (#3/tr) -- (#3/br) -- (#3/fr);

    \path[envelope, thin, gray] (#3/tl) -- (#3/tr); 
    
    \fill[letter] (#3/fl) -- (#3/eil) -- ++(90:\lheight) -| (#3/eir)
    -- (#3/fr) -- (#3/fl) -- cycle; 

  }

\newcommand{\jira}[3]{
  \def\ewidth{#2}
  \coordinate (#3/bl) at (#1); 
  \coordinate (#3/tr) at ($(#1) + (\ewidth,\ewidth)$); 

  \filldraw[jira] (#3/bl) rectangle (#3/tr);
}

\newcommand{\commit}[3]{ 
 \node[commit,minimum size=#2] (#3) at (#1) {};
}

\usepackage[textsize=tiny]{todonotes}

\newcommand{\eg}{\emph{e.g.}}
\newcommand{\ie}{\emph{i.e.}}
\newcommand{\etal}{\emph{et al.}\xspace}
\newcommand{\eref}[1]{Eq.~(\ref{#1})}
\newcommand{\Expect}[1]{\ensuremath{\mathbb{E}(#1)}}
\newcommand{\AMotif}{\ensuremath{|\text{AM}|}}
\newcommand{\Motif}{\ensuremath{|\text{M}|}}

\newcommand{\incfig}[1]{\includegraphics{img-gen/#1}}

\newcommand{\cl}{Conway's ``Law''\xspace}

\newcommand{\concept}{STMC\xspace} 
\newcommand{\measure}{dSTMC\xspace} 

\newcommand{\Camel}{\textsc{Camel}\xspace}
\newcommand{\Cassandra}{\textsc{Cassandra}\xspace}
\newcommand{\Groovy}{\textsc{Groovy}\xspace}
\newcommand{\HBase}{\textsc{HBase}\xspace}
\newcommand{\Spark}{\textsc{Spark}\xspace}
\newcommand{\TrafficServer}{\textsc{TrafficServer}\xspace}

\definecolor{colour1}{RGB}{202,0,32}
\definecolor{colour2}{RGB}{244,165,130}
\definecolor{colour3}{RGB}{146,197,222}
\definecolor{colour4}{RGB}{5,113,176}
\definecolor{colour5}{RGB}{200,200,200}
\tikzset{pstep/.style={draw,align=center,rounded corners,minimum width=5em,fill=white}}
\tikzset{ppath/.style={thick, -Stealth}}

\tikzset{envelope/.style={fill=white,draw,line join=miter,miter limit=2}}
\tikzset{letter/.style={envelope,fill=colour2}}

\tikzset{jira/.style={fill=colour2,thin,draw}}

\tikzset{commit/.style={fill=colour2,thin,draw,rectangle,rounded corners=2pt,inner sep=0pt, outer sep=0pt}}
\tikzset{nodeconn/.style={-Stealth,thin,draw}}

\tikzset{piclabel/.style={fill=black!40,inner sep=2pt,rounded 
    corners=1pt,text=white}}
\tikzset{marker/.style={fill=red,circle,outer sep=0pt, inner sep=0pt, minimum size=0.2em}}

\tikzset{backmark/.style={line width=1em,rounded corners=1pt,black!5}}
\newcommand{\manschgerl}{\textcolor{black!25}{\Huge\faMale}}
\newcommand{\glab}[1]{\footnotesize #1}
\newcommand*\circled[1]{\tikz[baseline=(char.base)]{
            \node[shape=circle,draw,inner sep=0.5pt] (char) {#1};}}

\makeatletter
\AtBeginDocument{\def\libertine@figurealign{}\libertineOsF}
\newcommand\libertineTabular{\def\libertine@figurealign{T}\libertineLF}
\makeatother

\begin{document}

\title{In Search of Socio-Technical Congruence:\\A Large-Scale Longitudinal Study}

\author{Wolfgang Mauerer,
  Mitchell Joblin,
  Damian A.~Tamburri,~\IEEEmembership{Member,~IEEE,}\\
  Carlos Paradis,~\IEEEmembership{Student Member,~IEEE,}
  Rick Kazman,~\IEEEmembership{Senior Member,~IEEE,}
  Sven Apel

  \IEEEcompsocitemizethanks{
    \IEEEcompsocthanksitem W.~Mauerer is with
    the Technical University of Applied Sciences\\ Regensburg, and Siemens
    AG, Corporate Research and Technology\protect\\
    eMail: wolfgang.mauerer@othr.de
    \IEEEcompsocthanksitem M.~Joblin is with Siemens AG, Corporate
    Research and Technology\\ and Saarland University\protect\\ 
    eMail: mitchell.joblin@siemens.com
    \IEEEcompsocthanksitem D.~A.~Tamburri is
    with the Technical University of Eindhoven\\ and the Jheronimus
    Academy of Data Science\protect\\
    eMail: d.a.tamburri@tue.nl
    \IEEEcompsocthanksitem C.~Paradis is with University of Hawaii\protect\\
    eMail: cvas@hawaii.edu
    \IEEEcompsocthanksitem R.~Kazman is with University of Hawaii\protect\\
    eMail: kazman@hawaii.edu
    \IEEEcompsocthanksitem S.~Apel is with Saarland University,
    Saarland Informatics Campus\protect\\
    eMail: apel@cs.uni-saarland.de}}

\markboth{Transactions on Software Engineering,~Vol.~XX, No.~XX, Month~20XX}%
{Mauerer \MakeLowercase{\textit{et al.}}}

\IEEEtitleabstractindextext{%
\begin{abstract}
  We report on a large-scale empirical study
  investigating the relevance of socio-technical congruence over key
  basic software quality metrics, namely, bugs and churn.  In particular, we explore whether alignment or misalignment of social communication
  structures and technical dependencies in large software projects influences software quality. To this end, we have defined a quantitative and operational notion of socio-technical congruence, which we call
  \emph{socio-technical motif congruence} (\concept). \concept is a measure
  of the degree to which developers working on the same file or on two related files, need to communicate.  As socio-technical congruence is a complex and multi-faceted phenomenon, the interpretability of
  the results is one of our main concerns, so we have employed a careful mixed-methods statistical analysis. In particular, we provide analyses with similar techniques as employed by seminal work in the field to ensure comparability of our results
  with the existing body of work. The major result of our study, based
  on an analysis of 25 large open-source projects, is that 
  \concept is \emph{not} related to project
  quality measures---software bugs and churn---in any temporal
  scenario. That is, we find no statistical relationship between the
  alignment of developer tasks and developer communications on the one
  hand, and project outcomes on the other hand. We conclude that, wherefore congruence does matter as literature shows, then its measurable effect lies elsewhere.
\end{abstract}

}
 
\maketitle


\section{Introduction}\label{sec:introduction}
The relationship between social and technical factors in software
engineering has received considerable attention in the past.  Despite
its importance,
to the best of our knowledge, a
systematic \emph{formalisation} and \emph{empirical evaluation} 
based on a rich set of longitudinal, triangulated
software engineering data is still lacking.

In this article, we describe a large-scale empirical study investigating
a formalisation of the well-known hypothesis of
socio-technical congruence. More specifically, we explore whether
alignment or misalignment of social communication structures and
technical dependencies in large software projects influences software
quality. To this end, we have defined a quantitative and interpretable
notion of socio-technical congruence, which we call
\emph{socio-technical motif congruence} (\concept). \concept is a
measure of the degree to which developers working on the same file or
on two related files, need to communicate.  As socio-technical
congruence is a complex and multi-faceted phenomenon, the
interpretability of the results is one of our main concerns, so we
have employed a careful mixed-methods statistical analysis. In
particular, we provide analyses with similar techniques as employed by
seminal work in the field~\cite{CataldoH13,Herbsleb:1999} to ensure
comparability of our results with the existing body of work.

The major result of our study, based on an analysis of 25 large
open-source projects from five different ecosystems, is that \concept
is \emph{not} correlated to basic and measurable project quality
outcomes---software bugs and churn---in any temporal scenario. That
is, we find no significant statistical relationship between the
alignment of developer tasks and developer communications on the one
hand, and project outcomes on the other hand.  We draw the conclusion
that if there is in fact a relation, it resides with other project
quality outcomes or at higher orders of organisational and technical
granularity. This conclusion is based on rigorous analyses of our
dataset, which spans hundreds of years worth of project data, ranging
over four different dimensions of software project and community
activity information (source code, mailing lists, 
issue-tracking logs, and commit logs and changes).

\begin{figure*}[htbp]
  \input{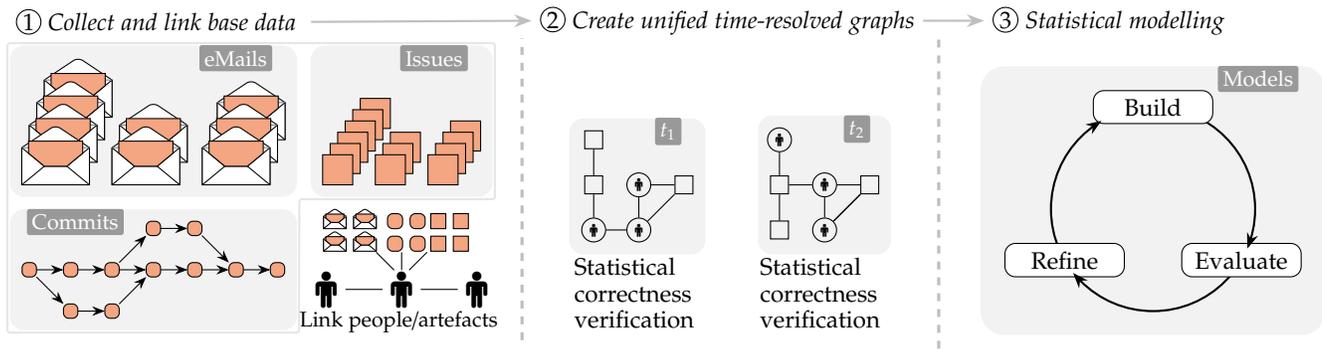}\vspace*{-1em}
  \caption{Overview of our research design.}
  \label{fig:overview_design}


  
  %
\end{figure*}



Although addressing a massive research corpus, our research design is
minimalistic to increase internal validity~\cite{SSA15}.  To this end
we: (a) formalise the concept of \concept in a simple, interpretable
manner, and we define a quantitative measure---\emph{degree of
  socio-technical motif congruence} (\measure)---of the
concept.  To ensure the generality and replicability of our results,
we analyze the ``residues''---the inevitable outputs and
by-products---of software development undertakings. The \emph{design}
that we analyze is provided by a project's code structure---files and
their relationships---and the \emph{communication structure} that we
analyzed is manifested by project members and their communication
relationships.  Together these structures comprise \emph{networks} of
technical artefacts and people. For the sake of generality, we chose
the simplest definition of a congruence pattern; (b) based on this
notion, we extract the two networks from software engineering records:
(i) code (\ie, analyzing files, their syntactic relationships, and
their co-evolution patterns) and (ii) people (\ie, project members and
their relationships, determined by analysing project mailing list and
issue-tracking data); (c) we capture elementary quality metrics:
code churn and bug density; (d) finally, we employ a
multitude of statistically interpretable techniques that qualitatively
and quantitatively establish the connection between \measure and
software quality.  Since looking for all imaginable consequences of
following or violating \concept is fundamentally impossible, we need to
pick specific measures, and thus we choose robust ones of practical
relevance.  We pay special attention not to focus on formal
statistical significance based on subjective
thresholds~\cite{McShane2019}, but center our analysis and discussion
on \emph{relevance}, together with a fully open and reproducible
approach.  An overview of the process that we follow is provided in
Figure~\ref{fig:overview_design}.

Our assumption is that, if \concept is relevant then the degree to
which technical dependencies match social communication structures
should affect project outcomes, which are measured as code quality
metrics.  In case of \emph{negative} findings, this of course leaves
the possibility that \concept is relevant for \emph{other} outcomes
that we do not consider in this study. In this article, we focus on
code quality metrics that are, at the same time: (a) most-immediately
impacting software stakeholders; (b) relating to organisational
structures according to the literature; (c) reflecting the lowest
level of abstraction of software projects and community activity. We
focus on software bugs~\cite{Aranda2009} and code
churn~\cite{MeneelyW12,KarusD12}, as discussed in
Section~\ref{sec:threats}. Note that, while there are many alternative
complexity metrics, these have led us to essentially identical results
and conclusions.  More precisely, we test: (a) whether there are
meaningful socio-technical patterns that arise from \concept; (b)
whether the presence of such patterns influences bugs and churn, (c)
how the impact of \measure compares to other influence factors, and
(d) whether effects induced by \measure endure over time.

Our results show that there is no observable relation between
\measure and bugs or churn; while this does not eliminate the
possibility of other beneficial effects of \concept on desirable
software qualities, our results exclude one major class of such
desiderata, and the methodology we employ can be re-used to 
examine such possibilities. Our results also demonstrate that the
quantitative influence of any possible effects caused by \concept on
quality outcomes is orders or magnitudes smaller than for other
influence factors, and is therefore not a worthwhile target for
substantial optimisation in practical industrial development.

The major contributions of this article are: 
\begin{compactenum}
\item[(1)] a robust and testable
\emph{quantitative} instantiation of \concept, together with a
measurable definition of its prevalence in real-world software
development projects; 
\item[(2)] a fully automatic, time-resolved analysis pipeline
  published as open source software\footnote{See
    \href{http://siemens.github.io/codeface/}{siemens.github.io/codeface/};
    a permanent archival copy is available at DOI
    \href{https://doi.org/10.5281/zenodo.4766404}{10.5281/zenodo.4766404}.}
    that combines heterogenous data sources and makes our study fully
    reproducible;
\item[(3)] a large-scale investigation using multiple statistical,
  interpretable and parsimonious approaches ranging from multivariate
  linear regression to elastic nets, of how \measure and software
  qualities are related, including a temporal analysis, based on
  formalisations of developer--artefact coupling and communication
  mechanisms;
\item[(4)] a discussion of the potentially wide-ranging impact of our findings on
common software architecture and folklore, in particular, related to
the crucial question of how to optimise human cooperation and
communication in development projects, and which aspects of human
cooperation reap the most benefits when supported and improved.
\end{compactenum}

All data generated by our study, the raw input data, and the scripts
to perform the required analytical computations, are available at the
supplementary web site
\href{https://cdn.lfdr.de/stmc}{cdn.lfdr.de/stmc} (an permanent
archive is stored at the DOI
\href{https://doi.org/10.5281/zenodo.4766388}{10.5281/zenodo.4766388}). The
website contains a comprehensive set of graphs and data that could not
be presented directly in the article.

\section{Related Work}\label{sec:related_work}
\begin{table*}[htb]
\scriptsize
\setlength{\tabcolsep}{1.4ex}
\caption{Overview of subject projects.}\label{tab:subject_projects}
  \centering 
 \libertineTabular\rowcolors{2}{gray!10}{white}
 \begin{tabular*}{\linewidth}{>{\scshape}lll 
   S[table-format=4.0]S[table-format=4.0]S[table-format=4.0]
   S[table-format=5.0]S[table-format=5.0]S[table-format=6.0]
   ccc}
  \toprule 
   \multicolumn{1}{l}{Project} & Lang & Domain & \multicolumn{3}{c}{\# Contributors} & {\# Commits} & {\# Issues} & {\# eMails}
           & \multicolumn{3}{c}{Analysis Period}\\
           \cmidrule{4-6}\cmidrule{10-12}
           &              &        &  {VCS} & {Issues} & {eMail}   &
           & {+ Comments} &        &  VCS   & Issues   & eMail \tabularnewline
   \midrule 
   \csname @input\endcsname {tab/tab.tex}
   \bottomrule
\end{tabular*}
\end{table*}


The relationships existing between organizational structure, its
characteristics, and the underlying relationships with software
structure and quality has been examined from several perspectives over
the years. On the one hand, the literature in software engineering has
most prominently focused on understanding and characterising the
relationships around socio-technical congruence
\cite{ValettoHECWW07}. In recent work, Kamola \cite{Kamola19}
investigates the effects of socio-technical congruence in the context
of multiple open-source software projects, with similar but deeper
investigations conducted by Syeed~\etal~\cite{SyeedH13}, but on a
single case study.  Similarly, Bailey~\etal~\cite{Bailey:2013} analyze
the effects of congruence over time and at larger scale, constituting
the theoretical basis for Betz~\etal~\cite{BSF+13}.  In
\emph{op.~cit.}, the authors provide a comprehensive overview of
socio-\hspace{0mm}technical congruence, relating to \cl and other
scientific literature pursuing an empirical perspective.  In much the
same timeframe, Cataldo~\etal~\cite{CataldoH13} provide evidence of
the impact implied in the aforementioned relations.  We seek to shed
light on the theoretical relations in the scope of the
\emph{organizational structure \(\leftrightarrow\) software structure}
congruence addressed by all the aforementioned work. Our attempt is to
dig into the organisational and technical macro- and micro-structures
\cite{sawyerartificial2003,ReinaMDT15} to find, characterise, and
possibly quantify existing empirical relations between, if any.

We do not assume any formulation of \concept as a theoretical basis;
rather, we seek to understand its measurable effect, if any at
all. The most relevant research related to our inquiry comes from
Colfer~\etal~\cite{colfer2010mirroring}, who formulate and investigate
the \emph{``mirroring hypothesis''}, that organizational structure
(represented as a network of co-\hspace{0mm}committing and
communicating developers) and software architecture (represented as a
design-structure matrix showing syntactic dependencies among software
components~\cite{dsm}) should be mirror images of each other.  The
authors do, in fact, find evidence supporting the mirroring
hypothesis. They consider eight open-source projects based on limited
sampling criteria, which brings some limitations to generalisability
of their results. However, despite considering the three-fold number
of projects, similar generalisability restrictions are still shared by
our results).  A similar objection can be made regarding the work by
Kwan~\etal~\cite{KwanCD12} as well as
Herbsleb~\etal~\cite{herbsleb1999beyondconwayslaw}, albeit we would
like to point ot that using an order of magnitude more subject
projects can not decisively solve the question of
generalizability. However, we feel that our work nonetheless provides
a substantial step forwards in terms of size and history of the
analysed software projects.  Other studies also investigate \concept
based on a qualitative discussion or using a smaller number of subject
projects, for instance
~\cite{Nagappan:2008,Syeed:2013,Herbsleb:1999}. Bird~\etal~\cite{Bird:2009}
successfully uses socio-technical networks to train predictive models
for build failures, but operates on a different level of artefact
granularity than our study, and uses an agglomeration without obvious
interpretation of graph measures (degree, centrality,\dots).  These
can additionally be different depending on the project under
consideration, since the mixture is driven by maximising prediction
accuracy. As we have discussed in Section~\ref{sec:rq2_modelling}, our
goal is to understand the influence of (given) interpretable,
operational characteristics of development efforts, and we do not see
a contradiction between our negative results and the positive results
given by Bird~\etal~\cite{Bird:2009}. It would be interesting to
cross-check their findings with our data, albeit we have to leave this
to future work.

Finally, from an organisational perspective,
it is already a known fact that the characteristics in an
organizational structure can affect the way in which software is
managed, operated, and evolved in that structure (\eg, see
Bird~\etal~\cite{BirdPDFD08}).
In the same vein, our study seeks to distill the relevant
structural characteristics that affect the organizational and
technical structures of open-source software projects; 
our long-term goal is
to construct a community quality model, through which both technical
and social debt \cite{Tamburri19,PalombaSZ17,TamburriKLV15} can be
assessed~\cite{Brown10debt,Kazman15debt}.  Much in the same way,
studies such as by Howinson~\cite{alonetogether} have tried to distill a
theory of socio-technical aspects in open-source projects (\eg,
motivation, coordination, or collaboration), but this and similar
approaches~\cite{MoonH14} fail to relate to concrete patterns and
metrics (e.g., collaborativity or cohesion across an organizational
structure) that could be used for planning preventive and corrective
actions. Conversely, from a different perspective, the same study of
the open-source phenomenon has led to several distinct formulations of
the same \emph{mirroring hypothesis}.

From a methodological perspective, we have chosen to analyze
structural congruence by identifying and studying the evolution of
``network motifs'', that is, recurrent patterns of socio-technical
relations that span the architectural and organizational structure, as
we describe in Section~\ref{sec:var_proc}.  The idea of using network
motifs to infer structural properties of networks is well-established
in many scientific fields~\cite{Alon2007,Hunsen:2020} but never before
seen in software engineering organizational research prior to
executing this work. Using network motifs to distill \concept mirrors
observational studies in open-source
communities~\cite{HowisonC14,MoonH16} that isolate positive
reinforcement patterns of organizational behavior and their impact on
software architecture.  In addition, network motifs have been widely
used in studying and understanding dynamic organizational and
socio-technical networks~\cite{GurukarRR15} that evolve over
time~\cite{ParanjapeBL16}, much like open-source software communities
and their software architectures. The work closest to ours at the
method level is limited to visualization of social relationship, for
example, Sarma~\etal~\cite{SarmaNH03}, most predominantly in the
context of global software engineering \cite{SarmaMWH09}.  We dig
deeper and wider to narrow down---by means of network-based motif
analysis---the patterns and recurrences of relationships claimed in
the literature.

\renewcommand{\cl}{Conway's Law\xspace}
\section{Research Design}\label{sec:research_design}
As we have illustrated in Figure~\ref{fig:overview_design}, the
research design of our study comprises multiple steps that range from
large-scale data collection and preprocessing from multiple data
sources via data fusion and validity verification to an iterative
model building and refinement process. We dedicate this chapter to
discussing variables and procedures used, starting with presenting our
research questions.

\subsection{Research Questions}\label{sec:research_questions}
In our study, we consider three research questions:

\begin{itemize}
\item\textbf{RQ1}---\emph{Is there a recurrent \concept pattern that
    enables us to quantify the amount of agreement between
    organisational and technical structures?}  To address this
  question we need to: (a) find basic patterns of relationships that
  an organised process exhibiting \concept may induce; (b) count and
  track the number of such basic patterns over time; (c) assess
  whether these observed patterns are due to random effects or not.

  Based on the counts, we define various measures for the
  \emph{degree of STMC} (\measure) to augment the notion of STMC with
  a quantified measure of how strong the concept is present in a given
  setting.
\item\textbf{RQ2}---\emph{Is \measure related with software quality?}
  If \measure is a meaningful concept, it must have a measurable,
  quantitative influence on software quality. That is to say, we would
  expect that some measures of software quality will vary with the
  degree of \concept exhibited by a project.
%
\item\textbf{RQ3}---\emph{Are there temporal implications of
    \concept?} One basic, recurring assumption about \concept, as
  outlined in Section~\ref{sec:related_work}, postulates that
  organisational and technical patterns in system development are in
  agreement, but there are two possible temporal implications by which
  any such effects can be considered; forward (advanced) and backward
  (retarded). This raises the questions of whether certain degrees of
  \measure at one point in time can lead to different properties of a
  project at a later point in time (and, if so, with what time delay)
  and whether measured project properties at one point in time
  influence communication structures at a later time?
\end{itemize}

\subsection{Variables and Procedure}\label{sec:var_proc}
We are interested in the connection between social and technical
aspects of software development projects. Socio-technical networks are
one of the standard tools used to abstract and represent how people
communicate and collaborate with each other and have been deployed in
many analysis
scenarios~\cite{LongS07,pinzger2008cds,begel2013social,Kazman16}.  Our
approach uses a network of people (developers) and artefacts (files)
that describes communication between people, dependencies between
artefacts, and interactions between persons and artefacts.

To obtain a reliable representation of a project's communication
patterns, we collect collaboration data on technical artefacts from
version control systems and data on communication from mailing lists
and issue trackers using established standard construction
methods~\cite{LongS07,bachelor2014,Ramsauer:2016} (details are in
Section~\ref{sec:network_construction}). Particular attention is given
to correctly resolving identities from different data sources~\cite{Joblin:2015}.

\subsubsection{Operationalising \concept}
The network structures that we construct allow us to bring the concept
of \concept into a precise, testable formulation through the use of
two network motifs. A \emph{network motif} is a sub-graph that is embedded in a larger graph.
Every socio-technical hypothesis needs to, in
some way or another, relate an artefact network (representing system
design via technical dependencies) to a developer network
(communication structure) by formulating \emph{constraints}.

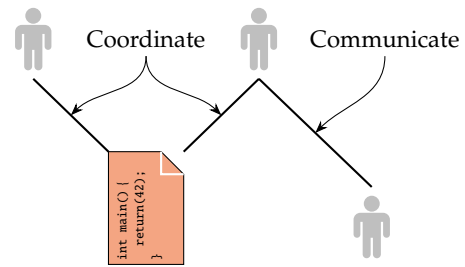
\begin{figure}[htbp]
  \begin{center}
    \def\earlength{0.3}
\def\codeheight{1.5}
\def\codewidth{1.0}
\begin{small}\begin{tikzpicture}
    \node[anchor=center] (p1) at (0.5,3) {\manschgerl};
    \node[anchor=center] (p2) at (3.5,3) {\manschgerl};
    \node[anchor=center] (p3) at (5,0.5) {\manschgerl};

    \coordinate (tr) at ($(1.5,0) + (\codewidth,\codeheight)$);
    \coordinate (tl) at ($(1.5,0) + (0,\codeheight)$);
    \coordinate (br) at ($(1.5,0) + (\codewidth,0)$);
    \coordinate (earl) at ($(tr) - (\earlength,0)$); 
    \coordinate (earr) at ($(tr) - (0,\earlength)$);
    \coordinate (ear) at (earl |- earr);
    \coordinate (pcent) at ($(p1)!0.5!(p2)$);
    
    \draw[fill=colour2] (1.5,0) -- ++(0,\codeheight) -- (earl) --
    (earr) -- (br) -- cycle;
    \draw[white,thick] (earl) |- (earr);

    \draw[thick] (tl) to (p1.south)
                 (tr) to (p2.south)
                 (p2.south) to (p3.north);

    \node[anchor=center,align=center] (coord) at (pcent) {Coordinate};
    \node[anchor=center,align=center,xshift=10em] (comm) at (pcent)
    {Communicate};

    \draw[thin] ($(p2.south)!0.5!(p3.north)$) edge[out=45,in=270,Stealth-] (comm);
    \draw[thin] ($(p1.south)!0.5!(tl)$) edge[out=45,in=270,Stealth-] (coord);
    \draw[thin] ($(p2.south)!0.5!(tr)$) edge[out=135,in=270,Stealth-] (coord);
    
    \node[anchor=north west,rotate=90,align=left] at (1.5,-0.05) {\tiny\texttt{int main()} \{\\[-0.4em]
      \tiny\texttt{\hspace*{1em}return(42);}\\[-0.4em]
      \tiny\texttt{\}}};
  \end{tikzpicture}\end{small}
  \end{center}\vspace*{-1em}
  \caption{Illustration of the organisational-silo effect; one side of
    the community is collaborating around a software artefact, but a
    communication silo exists around that
    artefact).}\label{fig:silo-effect}
\end{figure}

Our formulation of \concept reflects a recurrent, time-resolved,
sub-structure in a social network. In turn, a social network is a
weakly-typed graph \cite{Tamburri19} where multiple such time-resolved
sub-structures are possible. For example, consider the
\emph{organisational-silo effect} (see Fig.~\ref{fig:silo-effect}) or
even more complex \emph{community smells}, such as the
\emph{priggish members}, which would entail micro- and macro-structural
as well as sentiment-related social network motifs
\cite{prima-donna,TamburriKLV15,Dekker2014DeferringTE}.

Indeed, before delving into more complex motif analyses featuring
anti-patterns such as the effects outlined above, \concept reflects
one basic (general and thus fundamental) constraint. That is, when two
developers change a pair of files that have a dependency, these
developers may also need to communicate~\cite{Cataldo2008} (Note that,
since we are studying open source projects, we assume that most
developer communication occurs through formal project channels, \eg,
mailing lists and issue trackers, cf.~Sec.~\ref{sec:threats}).  This
communication constraint, for a pair of dependent artefacts, is
defined here as an \emph{indirect collaboration}. The collaboration is
indirect as it occurs through artefact dependencies, and is assumed to
be desirable. Conversely, if two developers collaborate indirectly and
do \emph{not} communicate, then this characterises an undesirable
situation: an \emph{anti-}motif pattern.

Figure~\ref{fig:motif_illustration} illustrates motif and anti-motifs
for indirect collaboration and non-collaboration. Since the four nodes
of the network can be conveniently be represented as a square, we
refer to indirect collaboration as a \emph{square motif}, and
non-collaboration as a \emph{square anti-motif}.

\newcommand{\setconwaygraph}{%
  \tikzset{x=1cm,y=1cm,nodes={draw, thick, black}}%
  \tikzstyle{person} = [circle, inner sep=0pt, minimum size=2mm]%
  \tikzstyle{artefact} = [rectangle, inner sep=0pt, minimum size=2mm]%
  \tikzstyle{communication} = [solid, thick]%
  \tikzstyle{modification} = [dashed, thick]%
  \tikzstyle{dependency} = [dotted, thick]%
  \tikzstyle{sg1} = [ultra thick, colour1]%
  \tikzstyle{sg2} = [ultra thick, colour2]}
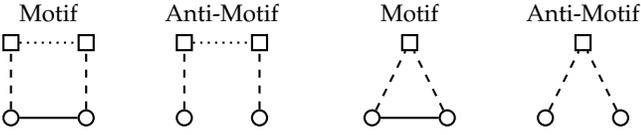
\begin{figure}[htbp]
  \setconwaygraph
  \begin{small}\begin{minipage}{0.23\linewidth}
    \begin{tikzpicture}
      \node[draw=white] at(0.5,1.4) { Motif };
      \node[person]   (p1) at (0,0) {};
      \node[person]   (p2) at (1,0) {};
      \node[artefact] (a1) at (0,1) {};
      \node[artefact] (a2) at (1,1) {};

      \draw[communication] (p1) -- (p2);
      \draw[modification]  (p1) -- (a1);
      \draw[modification]  (p2) -- (a2);
      \draw[dependency]    (a1) -- (a2);
    \end{tikzpicture}%
  \end{minipage}%
  \begin{minipage}{0.23\linewidth}
    \begin{tikzpicture}
      \node[draw=white] at(0.5,1.4) { Anti-Motif };
      \node[person]   (p1) at (0,0) {};
      \node[person]   (p2) at (1,0) {};
      \node[artefact] (a1) at (0,1) {};
      \node[artefact] (a2) at (1,1) {};
      
      \draw[modification]  (p1) -- (a1);
      \draw[modification]  (p2) -- (a2);
      \draw[dependency]    (a1) -- (a2);
    \end{tikzpicture}
  \end{minipage}\hfill%
  \begin{minipage}{0.23\linewidth}
    \begin{tikzpicture}
      \node[draw=white] at(0.5,1.4) { Motif };
      \node[person]   (p1) at (0,0) {};
      \node[person]   (p2) at (1,0) {};
      \node[artefact] (a1) at (0.5,1) {};

      \draw[communication] (p1) -- (p2);
      \draw[modification] (p1) -- (a1);
      \draw[modification] (p2) -- (a1);
  \end{tikzpicture}
  \end{minipage}%
  \begin{minipage}{0.23\linewidth}
    \begin{tikzpicture}[auto=left]
      \node[draw=white] at(0.5,1.4) { Anti-Motif };
      \node[person]   (p1) at (0,0) {};
      \node[person]   (p2) at (1,0) {};
      \node[artefact] (a1) at (0.5,1) {};

      \draw[modification] (p1) -- (a1);
      \draw[modification] (p2) -- (a1);
  \end{tikzpicture}
  \end{minipage}\end{small}
  \caption{Square (left) and triangle (right) motifs and anti-motifs
    measure the two most elementary forms of direct and indirect
    collaboration. Circles are developers, squares are
    artefacts. Solid edges indicate communication, dashed edges indicate modification, and
    dotted edges indicate dependency.}\label{fig:motif_illustration}
\end{figure}

Similarly we consider a second motif pattern that represents
\emph{direct collaboration}, which occurs when two developers modify a
\textit{single} artefact in the same time window.  According to the
ideas of \concept, they too should communicate.  Analogous to the
indirect collaboration, an anti-motif pattern occurs when the developers involved fail to
communicate.  As Figure~\ref{fig:motif_illustration} shows, the
resulting graph takes the form of a triangle.  Hence \emph{direct
  collaboration} will be referred to as a \emph{triangle motif} and
\emph{direct non-collaboration} will be referred to as a
\emph{triangle anti-motif}.

The square and triangle motifs and anti-motifs might be considered too
simplistic to represent non-trivial socio-technical effects.
Reformulating the organizational-silo effect defined above in terms of our
elementary motifs illustrates that, quite to the contrary, simpler
motifs can be seen as a ``basis'' for more involved positive or
negative socio-technical interactions. While positive triangle motifs are present in the context of such effects, so are also negative anti-motifs.
Consequently, presence of organisational-silo situations in the data implies a
dominance of negative over positive motifs, and our approach would
categorise the situation as adverse.  The example also highlights that
considering the proportion between positive and negative motifs is
important; this imbalance is, in fact, used in addressing RQ1.

\subsubsection{Socio-Technical Network Structure}
A socio-technical network is formalised by a
graph \(G=(V,E)\),
where the set of nodes \(V = V_{\text{d}}\cup V_{\text{a}}\)
comprises developers \(V_{\text{d}}\)
and technical artefacts \(V_{\text{a}}\),
source files in our case. The set of edges
\(E = E_{\text{comm}}\cup E_{\text{dep}} \cup E_{\text{mod}}\) 
models communication between developers by \(E_{\text{comm}}\subseteq
V_{\text{d}}\times V_{\text{d}}\) (solid lines in Figure~\ref{fig:motif_illustration}), modifications of artefacts by
developers via \(E_{\text{mod}} \subseteq V_{\text{d}}\times
V_{\text{a}}\) (dashed lines), and dependencies between artefacts
by \(E_{\text{dep}} \subseteq V_{\text{a}}\times V_{\text{a}}\) (dotted lines).

\subsubsection{Network Construction}\label{sec:network_construction}
We selected 25 open source projects listed in
Table~\vref{tab:subject_projects} from which to construct socio-technical 
networks. These projects vary in the following dimensions: (a) size
(lines of source code from 76~kLoC to over 1.1~MLoC, number of
developers from 25 to 1350), (b) age (time since first commit; three
to more than ten years), (c) programming language (we did place
attention on popular languages as determined by practical measures
such as the TIOBE index, but need to include the capabilities of the
various components of our analysis pipeline), (d) application
domain. We require (1) availability of public records for eMail and
issue tracking communication, and (2) availability of links between
version control system (VCS) commits and issues. We have additionally taken
practical importance and wide-spread deployment into account, albeit
the latter factors cannot be justified by entirely objective
criteria. Instead, we have also resorted to previous (subjective)
practical experience gathered in industrial projects. While this
selection procedure could have arrived at a different set of projects,
we do---owing to the large variation in the above dimensions---have
no reason to assume that the results would be drastically different
if we had chosen a different set of sample projects.

\paragraph{Gathering Raw Data}
To obtain the socio-technical network that describes communicative
relations between developers, technical relations between artefacts,
and modifications of artefacts by developers, we employed the tool
Codeface to analyze a project's version control system. Following
standard construction methods~\cite{Bird:2006}, an edge between an
artefact and a developer is present when the developer modifies an
artefact in a commit. We focus on files as artefacts.  Additionally,
entries in issues trackers are (as by the above requirement (2))
connected to commits, for instance by including an identifier of an
issue that is connected with a commit in the commit's description.
\emph{Comments} on issues link developers by a communication
relationship. Taken together, this establishes a mapping between
artefacts (touched by the commit) and issues, and permits
identification of communication over a subset of files for a given
time period. A developer network is then a graph where each vertex is
a developer, and each weighted edge is the sum of comments between
pairs of developers across all issues.

\paragraph{Inferring Communication, Dependencies and Qualities}
Defining what constitutes developer communication, artefact
dependencies and software quality are of course subject to expectations
and requirements. The literature employs many different conventions,
and we support multiple construction approaches to increase the
generality of our considerations (note that we provide more
technical details on the collection of base data and network
construction in the appendix):

\begin{itemize}
\item \emph{Artefact--artefact dependencies:} We denote a dependency between
  a pair of artefacts if they have been involved in commits together, if they
  have static (language-level) dependencies, such as calling or inheritance
  relationships, or if they are semantically (textual content of
  source code comments) related. See the appendix for details.
\item \emph{Developer communication:} We capture communication
  relationships from emails (messages and responses), issue trackers
  (entries and comments), and a combination of both.
\item \emph{Quality indicators:} We use the number of bugs and amount
  of churn (committed lines of code) per file as simple, robust and easily
  quantifiable measures of software quality. From the bug count
  per file for a give temporal range, we compute the bug density by
  normalising the bug count by the size of the file as given by the
  number of lines it contains.
\end{itemize}

Our analysis is performed on all \(3\times 3\times 2=18\) resulting 
combinations of dependencies, communication mechanisms, and quality indicators.

The construction methods for artefact-artefact dependencies
 are determined as follows:

\begin{enumerate}
\item \emph{Static dependencies}~\cite{Browning16} between artefacts (such as function calls)
  are obtained using the Understand\footnote{https://scitools.com}
  tool. A dependency structure matrix (DSM) with one row and column for
  each artefact captures dependencies from one artefact to
  another. This coupling mechanism represents the most basic structure of a
  software system.
\item \emph{Evolutionary dependencies}
  (co-changes)~\cite{WongCVSS09,KhosraviC09} are obtained using
  Codeface.  An edge between artefacts arises when
  two artefacts are jointly modified in one commit. 
\item Artefact--artefact respectively semantic coupling (attempting to
  capture the developers' mental model of a system) is obtained by
  semantically analysing comments~\cite{Sangal2005} respective key
  terms associated with source code~\cite{6976091,Joblin:2016}. Latent
  semantic indexing techniques are employed to identify relationships
  between key terms, which are then aggregated at the file level and
  interpreted as \emph{semantic dependencies} between artefacts. These
  dependencies are also obtained using the Codeface tool.  This
  coupling mechanism focuses on coordination and implicit aspects of
  the software architecture.
\end{enumerate}

Developers communicate using various \emph{communication
  channels}. Each channel may reflect a different interpretation of
the developer's network~\cite{Iqbal:2013}, even if the construction
method is the same. We have analyzed communication from both
\textit{issue trackers} and \emph{mailing lists} because these
resources are available for a wide range of projects, and are known to
provide reasonably reliable and comprehensive information on developer
communication~\cite{6976091,Meneelymeasurements}. Note that, while our
analysis pipeline produces weighted networks, where edge weights
represent communication frequency (and related measures, depending on
the meaning of an edge), we deliberately ignore weights. To the best
of our knowledge, we are not aware of any objectively justified
cut-off value below which edges should not be considered because of
their irrelevance. Additionally, introducing the concept of relevance
would entail further challenges: A pair of developers could frequently
debate unimportant trivia, which would nonetheless lead to a heavily
weighted communication edge. Another pair could engage in a short
discussion on issues of the gravest magnitude, which would yet deliver
a light-weight edge. We are not aware of any means to reliably and
objectively weight the content of communication, or provide related
means for other edge types.

To measure software quality, we count the number
of bugs associated with an artefact (bug density) and compute
the magnitude of churn (changed lines per artefact).  Clearly,
an increasing number of bugs is equivalent to decreasing software
quality. Likewise, a high change rate (churn) over extended periods of
time on a given artefact is indicative of low software
quality~\cite{MeneelyW12}.

Since properties of projects are not static but change over time,
 we extract the previously described
co-variables not just for single static code snapshots, but we construct
a series of networks \((G_{0}, G_{1}, \dots, G_{n})\) where network
\(G_{i}\) includes activities that took place
during the temporal interval \([i, i+\Delta t]\) (\(i=0\) denotes the
first point in time for which all data sources provide artefacts). We
uniformly use a window size \(\Delta t\) of three months. Meneely and
Williams~\cite{Meneelymeasurements} have shown that the effect of
enlarging the window beyond three months is marginal for a wide range
of analysis tasks; we discuss in Section~\ref{sec:sensitivity}
that their argumentation also holds for our analyses.

\subsubsection{Pattern Detection}\label{sec:network_pattern_detection}
\begin{figure}[htbp]
  \setconwaygraph
  \tikzset{x=1cm,y=0.7cm,nodes={draw, thick, black}}%
  \vbox{\hbox{\begin{tikzpicture}
  \node[artefact, sg1] (a40) at (0, 4) {};
  \node[artefact, sg1] (a41) at (2-0.5, 4+0.3) {};

  \node[person, sg1]   (p30) at (1-0.4, 3) {};
  \node[person]   (p31) at (2-0.5, 3+0.2) {};
  \node[person]   (p32) at (3, 3) {};

  \node[person, sg1]   (p20) at (0, 2) {};
  \node[artefact] (a21) at (1-0.1, 2+0.2) {};
  \node[artefact, sg2] (a22) at (2-0.2, 2) {};
  \node[artefact] (a23) at (3+0.2, 2+0.1) {};

  \node[person]   (p10) at (1+0.4, 1) {};
  \node[person, sg2]   (p11) at (2+0.5, 1) {};

  \node[person]   (p00) at (0, 0+0.3) {};
  \node[artefact] (a01) at (1, 0) {};
  \node[person, sg2]   (p02) at (2, 0) {};
  \node[artefact] (a03) at (3+0.2, 0+0.3) {};


  \draw[communication] (p00) -- (p10);
  \draw[communication] (p00) -- (p20);
  \draw[communication, sg2] (p02) -- (p11);
  \draw[dependency] (a01) -- (a21);
  \draw[dependency] (a03) -- (a23);
  \draw[modification, sg2] (p02) -- (a22);

  \draw[modification, sg2] (p11) -- (a22);
  \draw[communication] (p10) -- (p31);

  \draw[dependency] (a22) -- (a23);
  \draw[modification] (a21) -- (p31);
  \draw[modification] (a22) -- (p31);

  \draw[communication] (p31) -- (p32);
  \draw[modification]  (p31) -- (a41);

  \draw[dependency, sg1]  (a40) -- (a41);
  \draw[modification, sg1]  (p20) -- (a40);
  \draw[modification, sg1]  (p30) -- (a41);
\end{tikzpicture}
\hspace*{3em}
\newcommand{\partnodedown}[5]{
  \begin{scope}
    \clip (#2,#3) -- (#2+0.3,#3) -- (#2+0.3,#3-0.3) -- (#2-0.3,#3-0.3) -- (#2-0.3,#3) -- (#2,#3);
    \node[#5, #4] (#1) at (#2, #3) {};
  \end{scope}
}
\newcommand{\partnodeup}[5]{
  \begin{scope}
    \clip (#2,#3) -- (#2+0.3,#3) -- (#2+0.3,#3+0.3) -- (#2-0.3,#3+0.3) -- (#2-0.3,#3) -- (#2,#3);
    \node[#5, #4] (#1) at (#2, #3) {};
  \end{scope}
}
\begin{tikzpicture}
  \partnodeup{p30}{0}{3}{sg2}{person}
  \partnodedown{p30red}{0}{3}{sg1}{person}

  \partnodeup{a31}{1+0.5}{3+0.2}{sg2}{artefact}
  \partnodedown{a31red}{1+0.5}{3+0.2}{sg1}{artefact}

  \node[artefact] (a32) at (2+0.5, 3+0.2) {};

  \node[person]   (p20) at (0-0.3, 2) {};
  \node[person, sg2]   (p21) at (1, 2) {};
  \node[artefact] (a22) at (2-0.4, 2-0.1) {};

  \node[artefact] (a10) at (0+0.2, 1+0.2) {};
  \node[person]   (p11) at (1, 1) {};
  \node[person, sg1]   (p12) at (2+0.3, 1) {};

  \node[person] (p00) at (0, 0+0.3) {};
  \node[person] (p01) at (1, 0+0.1) {};
  \node[person] (p02) at (2, 0) {};


  \draw[communication] (p01) -- (p12);
  \draw[communication] (p01) -- (p11);
  \draw[communication] (p02) -- (p12);
  \draw[modification]  (p01) -- (a10);

  \draw[dependency] (a10) -- (a22);
  \draw[communication] (p11) -- (p21);

  \draw[communication] (p20) -- (p21);
  \draw[communication] (p20) -- (p30);
  \draw[communication, sg2] (p21) -- (p30);
  \draw[modification, sg2] (p21) -- (a31);
  \draw[dependency] (a22) -- (a31);
  \draw[dependency] (a22) -- (a32);

  \draw[modification, sg1]  (p30) -- (a31);
  \draw[modification, sg2, dash phase=4.5pt]  (p30) -- (a31);
  \draw[modification]  (p12) -- (a32);
  \draw[modification, sg1]  (p12) -- (a31);
\end{tikzpicture}

}
    \vspace*{2mm}
    \hbox{\begin{small}\begin{tikzpicture}
          \draw[-Stealth] (0,0) -- (8,0); 
          \draw (2,-0.25) -- (2, 0.25) { };
          \draw (6,-0.25) -- (6, 0.25) { };
          \node[draw=none] at (8.5, 0) { \(t\) };
          \node[draw=none] at (2, -1) { \(G_{t_{1}}=(V_{1}, E_{1})\) };
          \node[draw=none] at (6, -1) { \(G_{t_{2}}=(V_{2}, E_{2})\) };
        \end{tikzpicture}\end{small}}}\vspace*{-1em}
    \caption{Detecting \concept patterns in time-resolved
      collaboration graphs. Nodes and edges have the same meaning as
      in Figure~\ref{fig:motif_illustration}; sub-graphs that
      correspond to motifs are emphasised using non-black colours and
      thicker strokes.}\label{fig:detect_motifs}
\end{figure}
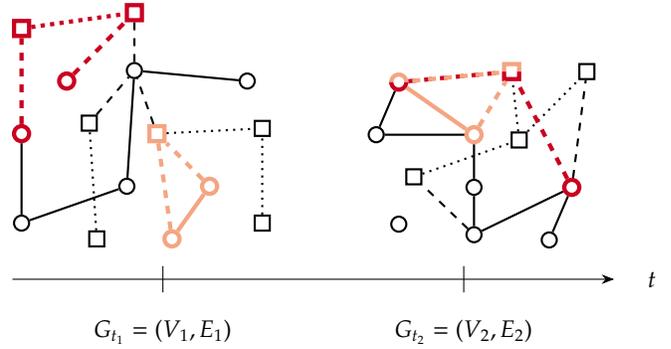

We have introduced \concept informally. We will now
formalise this concept.  Given a two-mode graph 
\(G=(V,E)\) that models the artefact-developer network,
and a motif described by another two-mode graph \(M = (V', E')\),
we need to count how many sub-graphs
\(\tilde{G}=(\tilde{V}, \tilde{E}) \subseteq G\)
with \(\tilde{V}\subseteq V\)
and \(\tilde{E}\subseteq \tilde{V}\times \tilde{V}\)
are isomorphic to \(M\),
respectively if there exists a function \(f: V_{0}\mapsto V'\)
such that
\((v_{0}, v_{1}) \in E_{0} \Rightarrow (f(v_{0}), f(v_{1}))\in E'\).
This is similar to the well-known (counting variant of the) sub-graph
isomorphism problem.  However, we need make sure that \(f\)
is \emph{injective} to ensure that missing edges in the motif are
mapped to missing edges in the larger graph (otherwise, any motif in
the graph would also be counted as an anti-motif).  This is known as
induced sub-graph matching, and has the added computational benefit 
that the decision variant is in P (unlike the general problem that is
known to be NP-complete), which makes our approach practically
tractable. The approach is illustrated in
Figure~\ref{fig:detect_motifs}.

\begin{figure}[htbp]
  \incfig{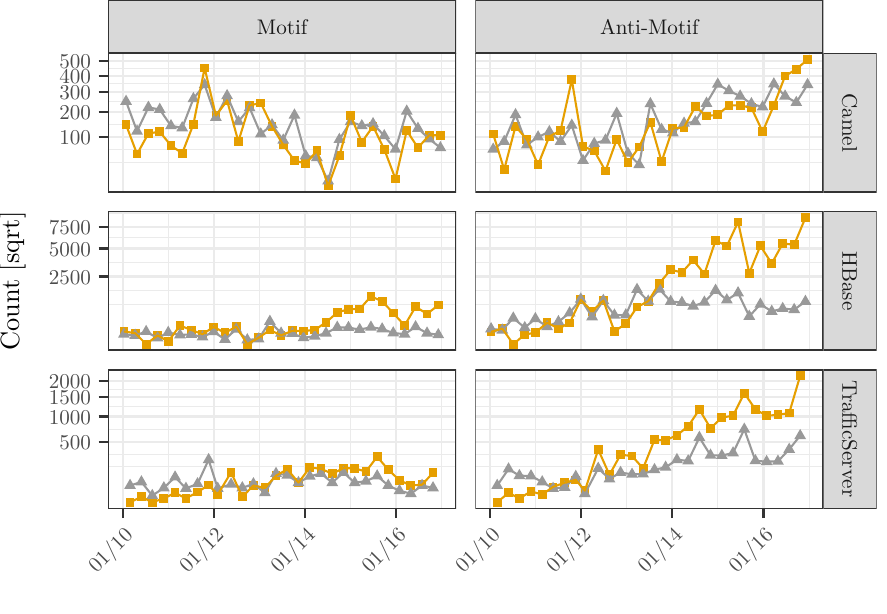}\vspace*{-1em}
  \caption{Typical time-resolved square and triangle motif and
    anti-motif counts for representative sample projects. Triangles
    and squares in the graph represent the according motifs and
    anti-motifs.}\label{fig:motif_ts_abs}
\end{figure}

We obtain four measurements of motif patterns for every time window
per project: (1) square motif counts, (2) square anti-motif counts,
(3) triangle motif counts, and (4) triangle anti-motif counts.  Since
motifs characterise \emph{agreement} with \concept, and anti-motifs
\emph{disagreement} with \concept, and they are measured on each time
window, we are able to measure \concept over a project's lifetime
through \emph{direct} (triangle motif) and \emph{indirect} (square
motif) collaboration. In particular, we can determine the extent of
\concept in a project by computing the ratios between the occurrences
of motifs and anti-motifs.  Figure~\ref{fig:motif_ts_abs} illustrates
the magnitudes of motifs that occur in some of the subject projects;
while we carefully establish the validity of the measurements and
their relation to the real world in Section~\ref{sec:analysis:rq1}, we
observe that the triangle and square motifs and anti-motifs develop
similarly over time, and do not fluctuate
randomly. The large absolute numbers of motif counts (hundreds to
thousands for each time window) underline that statistically relevant
findings are accompanied by an appropriate effect size. The congruent
temporal development of triangle and square motif and anti-motif
counts, which is visually apparent, serves as intuitive sanity check that
the measurements are consistent.

\paragraph{Configuration Model Hypothesis Testing}\label{sec:config_model}
To address RQ1, any measurement of \measure using the four motif
patterns above must be accompanied by an empirical evaluation of
whether the counted motifs are occurring randomly, given the size of
the network and the chance that any pair of nodes will or will not be
connected randomly.  To verify this, we chose to employ a
\emph{configuration model}---a generalised random graph
model~\cite{Newman:2010} that allows for constructing a multitude of
graphs with the (fixed) degree sequence of an empirical reference
graph, while randomizing all other structure.  Given a degree
sequence---the monotonic, non-increasing sequence of vertex degrees;
recall that many topologically different graphs can have the same
degree sequence, which is at the heart of the idea---of a network,
the model \emph{randomly} chooses, for every node, which other node
it will be paired with (bounded by the specified node degree), while
maintaining the overall degree sequence. To construct a graph
variant, the algorithm first removes all edges from the original
graph, and assigns the elements of the degree sequence to the nodes,
encoded in half-edges. Then, two such half-edges are chosen
uniformly at random, and connected to a complete edge.  
Iteratively, another pair from the remaining half-edges is chosen and
connected, until there are no half-edges left (we omit details on how
to handle networks with an uneven number of nodes).

Since we deal with a two-mode network that contains two
different types of nodes, we need to additionally make sure that the
functional characteristics of the nodes (besides their number) are
preserved. The approach employed by our calculations relies on methods
initially invented for structurally related biological
problems~\cite{Gobbi2014,IoroXX}, and is known to maximise
dissimilarity between source and rewired graph (as measured by the
Jaccard Index~\cite{jaccard1901etude}), while minimising the number of
transformation (edge switching) steps that need to be performed.

We iteratively perform the rewiring process \(N\) times for each time
window \(t\), determine the number of triangle and square
\hbox{(anti-)} motifs (indexed by \(m\)) in each rewired configuration
model graph, and obtain a (discrete) probability distribution
\(p_{m,t}(n) = \hat{c}_{m,t}/N\), where \(\hat{c}_{m,t}\) denotes the
count for a motif \(m\) in time window \(t\) (\(N\) is chosen
sufficiently large to guarantee convergent
results). Providing statistical certainty requires
  considerable computational effort (measured in CPU months), which we
  would like to highlight: While this may be a purely technical issue
  from a conceptual point of view, it poses a considerable amount of
  practical challenges that can only be solved by using
  state-of-the-art methodology from distributed computing and big data
  analysis which, in turn, is possible because in a completely
  reproducible setting thanks to the efforts of researchers in the
  statistical, numerical and high performance computing
  domains~\cite{Tang2016,Venables2002,Kuznetsova2017,Wood2015,Wood2011b,
    Wood2003,Wood2017,Wood2004,Wood2011,Wickham2007,Wickham2011,Wickham2016,R2018}.

We can then relate this simulated distribution to the number \(c_{m,t}\)
of (anti-) motifs \(m\) observed in the measured, real network at time
window \(t\), and employ standard statistical techniques as in~\cite{Joblin:2015}
to compute how probable it is to obtain the empirically observed count in a
socio-technical network that arose randomly, just like the rewired
graphs. If the real-word data turn out to be highly improbable, we
conclude that the observed counts do not stem from a random process,
but must arise from a meaningful and intentional interaction of
developers (of course keeping the restrictions of statistical
hypothesis testing in mind). The validity of our approach is further
discussed in the appendix on page~\pageref{sec:validity_base_data}.

To establish a bridge between appearance counts of the four motif
patterns in given time window and statistically significant effects on
measurable project quality or other outcomes, we propose an
\emph{artefact participation} measure. We count, for \emph{each
  artefact}, how often the artefact occurs (participates) in each one
of the counted motifs. These numbers can then be visualised and
analysed in a multitude of ways, for instance using time series
methods, or, more importantly, by correlating them with various
quality observables.

\paragraph{Effect of \measure on Software Quality}
We have focused our investigation on the consequences of \concept
on \textit{issues} (bugs and problems identified
in a project's issue-tracking system) and \textit{churn} (the number
of lines modified, over time, as determined by a project's
configuration management system), as these are commonly used measures
of project quality. If \concept is present in a project, or in some
development time windows, it should have a measurable influence on
project quality.  We test this by correlating \measure with the
aforementioned quality measures. Besides using quality and
communication data from the same time window, we also check whether there
are any temporally distributed effects (changes in
projects take time to manifest; for instance, changes in \measure in
one time window may lead to better software quality in a following
window).

\section{Research Question 1 -- STMC Patterns}\label{sec:analysis:rq1}
After all formal definitions and technical procedures have been
introduced, let us commence to discussing research question~1, which
concerns the identification of statistically valid and quantifiable
\concept patterns.

\subsection{Network Construction}
We base the investigation of the first research question on our
operationalisation of \concept. For each motif, analysis, and time
window, the number of motifs present in the real data is determined by
an induced sub-graph isomorphism calculation on the socio-technical
collaboration graph, and then compared to the counts obtained for the
rewired graphs. This is illustrated in Figure~\ref{fig:null_model_ts}
for a subset of the time ranges for project \HBase (with co-change as
dependency mechanism, eMail as communication mechanism, and Jira as
bug tracking mechanism), shown for square motif and anti-motif. Owing
to the large magnitude of information that arises from the many
projects, analyses, and time ranges that we consider, we
can discuss only representative examples. The reader can refer to the
accompanying website for the complete data sets and analyses.

\emph{Time-resolved analysis}: For each motif, analysis, and time
window, we perform a one-sample t-test resulting in a p-value with the
Null hypothesis that the empirical (\ie, real-world) sub-graph count
is compatible with the data obtained from graph rewiring. We want to
emphasise two aspects: Firstly, unlike fishing for significant
results, the \emph{same} hypothesis is tested on different
analyses.  
Secondly, our multiple testing approach cannot unintentionally
generate a small number of significant results from a set of otherwise
insignificant results by just increasing the sample size of the number
of experiments~\cite{Reinhart2015}. In contrast, our tests lead to
\emph{rejection} in almost all instances. Consequently, we are not
affected by \emph{false discovery rate}-type~\cite{Benjamini1995}
issues.
  The repeated tests result
in a distribution of p-values, most of which are extremely small,
below \(10^{-2}\). The full distribution is available in the
online supplement.

We note that p-values tend, on average, to be slightly larger for the
triangle than for the square motif, which can intuitively be easily
understood as the simple triangle motif is more likely to appear
randomly in a graph structure than the structurally more complicated
square motif.

\begin{figure}[htbp]
  \incfig{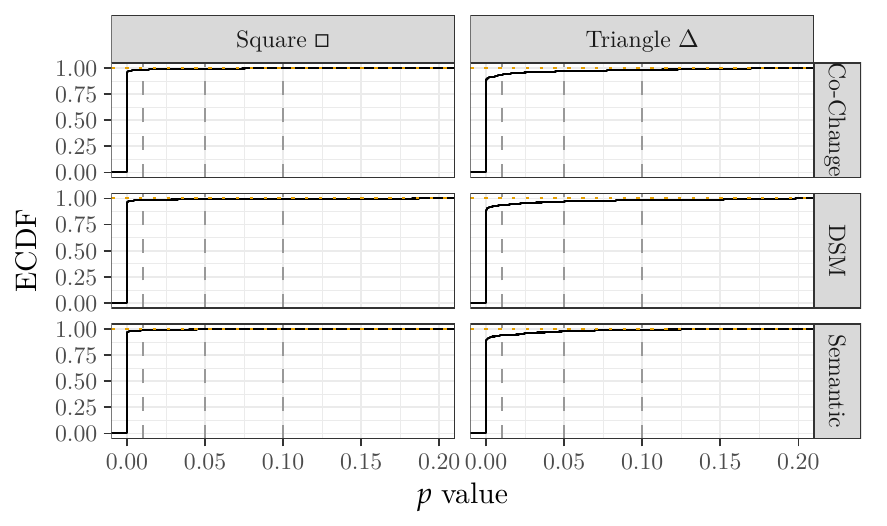}\vspace*{-1.5em}
  \caption{Empirical cumulative distribution function (ECDF) for p values
    obtained with time-resolved statistical significance testing for
    network accuracy, resolved by motif type (horizontal panels) and
    coupling mechanisms (vertical panels). The range is reduced to a
    maximal value of 20\% because no appreciable changes happen in
    larger regions.\newline
    Vertical, dashed lines indicate the commonly employed significance
    levels of 1\%, 5\% and 10\%, but these are only used to guide the
    eye. The horizontal dotted line illustrates the 100\%
    level of the ECDF.}\label{fig:p_value_dist}
\end{figure}

\emph{Global analysis:} To gain a data-set--wide overview, we show the
distributions that arise for the individual projects, coupling
mechanisms and motif types in Figure~\ref{fig:p_value_dist}.  We can
reject \(H_{0}\) at an essentially arbitrary level in the overwhelming
majority of cases for all analysis combinations, indicating that the
features we detect in the data are highly non-random in nature, and
independent of the projects that we analyse.

While the figure shows guiding lines for typical significance levels,
these are only used for \emph{illustrative} purposes, and should not
be taken as a pre-specified significance level~\cite{Drummond2016} on
part of the authors. Following the latest recommendation of the
statistics community~\cite{Wasserstein2019,Wasserstein2016}, we also
do not make any claims about significance or insignificance of our
(repeated) measurement results. Our major claim is that it is highly
unlikely that the observed socio-technical structures stem from random
processes, and that the amount of chance required to obtain the results
in a random scenario makes, together with the confirmatory
sociological investigation in previous work~\cite{Joblin:2015}, a
compelling argument for the meaningfulness of our base network
construction.



\subsection{Quantitative Measures for the Degree of \concept}
With the non-random occurrence of the two types of motifs established, and
for various combinations of technical dependencies and communication
channels, defining \emph{\concept} is a matter of relating the two
quantities (motifs and qualities) in an appropriate way. To ease interpretability, the
quantity should be \emph{normalised}, and should not depend on the
absolute motif \emph{counts}. However, since this implicitly requires
\concept to be scale invariant with respect to artefact size, which we
do not want to mandate a priori, we also allow the size of artefacts
to influences a second, complementary measure of \concept. From the
many possible mathematical realisations of these requirements, we have
chosen \emph{signed motif percent difference} \(r(\AMotif, \Motif)\)
(with \Motif{} and \AMotif{} denoting the number of motifs and
anti-motifs, respectively) and the \emph{LoC normalised motif
  difference} \(l(\AMotif, \Motif)\) given by
\begin{align}
  r(\AMotif, \Motif) &\coloneqq 2\frac{\AMotif-\Motif}{\AMotif+\Motif},\\
  l_{\text{a}}(\AMotif, \Motif) &\coloneqq \frac{\AMotif-\Motif}{|\text{a}|},
\end{align}
which result in a normalised quantity \(r\in [-2,2]\), and an
unrestricted quantity \(l \in (-\infty, \infty)\) that depends on the
size of the artefact given by \(|\text{a}|\)---for the case of
  files as artefacts, which we use in our study, \(|\text{a}|\) is
  given by the number of source lines.  \(r\) quantifies the relative
difference between two measures \(a\) and \(b\) without having to set
one as base quantity. Instead, \(r\) computes both, the relative
difference between \(a\), \(b\) and \(b\), \(a\), and then averages the
results. Note that the covariates could also be related via, for instance,
  \(\Motif/(\AMotif+\Motif)\), but this does not change any of the
  interpretations presented in the article. We found that the chosen
  ratios provide, in our opinion, the fewest mathematical surprises
  that require explaining.

Another property of \(r\) that makes it well suited to describe
socio-technical aspects is that it exhibits less rapid
variations and fewer discontinuous changes than the constituent
quantities \(\Motif\) and \(\AMotif\). Social processes can reasonably be
expected to only change gradually when a project evolves, and so
should any quantities describing the associated phenomena.






To interpret \(r\), consider three cases. When the number of
(positive) motifs is constant (\(\Motif=c=\text{const}\)), and the
number of (negative) anti-motifs grows,
\(\lim_{\AMotif\rightarrow\infty} r(c, \AMotif)=2\).  A growing number
of anti-motifs relative to a given number of motifs means
a \emph{decreasing} \measure; the quantity \(r\) approaches \(2\) in
this case.  Contrariwise, with an \emph{increasing} \measure, that is,
\(\AMotif=c=\text{const}\) and a growing number of (positive) motifs,
it holds that \(\lim_{\Motif\rightarrow\infty} r(\Motif, c)=-2\). When
positive and negative motif counts balance each other
(\(\AMotif=\Motif\)), then \(r(c, c)=0\) (to handle a
  pathological case, we set \(r(0,0) \coloneqq 0.\)). The measure
describes two aspects: Sign indicates a regime of \concept (-1) and
of anti-\concept (+1), and magnitude provides a normalised effect size.

Interpreting \(l_{\text{a}}(\AMotif, \Motif)\) is similar, except that
increasing unbounded differences between motif- and anti-motif counts
lead to increasing, unbounded values of \(l\). Most importantly, the
sign behaviour of \(l_{\text{a}}(\cdot, \cdot)\) is identical to
\(r(\cdot, \cdot)\). The convention has been chosen with the form of
regression models \(y \propto \beta\cdot r(\AMotif, \Motif)\) in mind:
The left-hand sides considered in this study are bug density and
churn; both are ``bad'' (negative connotation) when they are high, and
``good'' (positive connotation) when they are low. When there are more
positive motifs than (negative) anti-motifs, both \(r\) and
\(l_{\text{a}}\) are negative.  Consider the case in which the
coefficient \(\beta\) is positive. Then, the regressand \(y\)
\emph{decreases} (which is desirable, since a smaller bug density is
``good'') when the absolute magnitude of \(r\) and \(l_{\text{a}}\)
\emph{increases}. Consequently, under the given sign convention,
positive regression coefficients \(\beta\) represent a ``good'',
desirable scenario, while negative coefficients represent a ``bad''
scenario. This is supposed to ease interpreting the result graphs that
follow.

We do not implicitly assume that one or the other of \(r\) or
\(l_{\text{a}}\) is more ``natural'' or preferable in any way;
both indeed describe different aspects of the problems. We leave the
decision to the data and their analysis.

\subsection{Answering Research Question 1}
Considering construction and validity assurance of our socio-technical
networks obtained from real data, we have established that motif and
anti-motif patterns occur strongly non-randomly in collaboration graphs.
They vary moderately, yet distinctly over time, and are therefore are
not just a global static property of a project. Since the chosen
motifs relate key social and technical aspects of a project, and
capture any changes of these relations over time, they serve as the
desired indicators for \measure, in particular given the measures
\(r\) and \(l\) that relatively relate counts of the two quantities.
Therefore, we answer RQ1 \emph{affirmatively}.

\section{Research Question 2---Relation to SW Quality}
As we have outlined earlier, any meaningful hypothesis regarding
\concept must clearly indicate the observable (positive) consequences
of high \measure, and must similarly indicate the (negative)
consequences of lower values of \measure.  We are now interested in
studying the influence of \concept on software quality, as
characterized by bug density and churn. Assuming there is a meaningful
relationship between social and technical aspects of software
development as captures by \concept, then a higher \measure value
shall lead to fewer bugs (because developers are communicating
``appropriately''), and a lower \measure shall lead to more bugs. The
available data allow us to test this relationship by considering how
\measure of a given artefact (determined by the positive and negative
motifs the artefact participates in), and software quality (number of
bugs, churn) associated with the artefact are related. We investigate this
relationship in a multi-stage process that progresses from an in-depth
analysis of specific projects with various statistical techniques to a
more general, broader-scale investigation that confirms the detail
findings for a large number of projects, and for long temporal
histories.

\subsection{Regression Modelling}\label{sec:rq2_modelling}
Understanding how a set of measured variables influences a quantity of
interest, and separating the effect of one particular variable from
the effect of the remaining variables, is a problem that has been
comprehensively considered in the statistical
literature~\cite{Hastie2009, Fahrmeir2013}. Previous work has
considered multivariate linear models (\eg,~\cite{CataldoH13}) or more
advanced forms of regression for this purpose to understand the nature
of socio-technical effects in software engineering. We perform a
three-stage approach of increasing sophistication and generality:

\begin{compactitem}
\item[(1)] We compute multivariate linear and logistic regression models on
  our data sets, and carefully evaluate validity, significance, and
  effect size~\cite{Greenland2016}. We conclude that they
  are not able to provide satisfactory evidence for strong relations
  between \measure and software quality. Additionally, testing the
  fulfilment of modelling preconditions shows that the class of
  analysis techniques is not the optimal choice for our data. We also
  discuss that we cannot reproduce earlier results on socio-technical
  relationships based on slightly different notions of \measure.
\item[(2)] To determine how deficiencies of the straightforward models are
  related to how the data are transformed, we employ generalised
  additive models that introduce additionally required non-linear
  transformations of predictors in a non-parametric way, yet keep
  (compared to many machine learning 
  approaches~\cite{Breiman2001}) the analysis outcome
  interpretable. The compatibility of the improved models with the
  given data are en par with previous approaches, but still strongly
  support our conclusion that no meaningful relationship exists
  between the \measure and the considered quality properties. In particular, the
  (automatically chosen) non-linear transformations only increase the
  importance non-\concept predictors, confirming that deficiencies of
  simpler modelling approaches do not cause the non-relevance of
  \concept predictors.
\item[(3)] Ensuring correct models in terms of satisfying mathematical pre-
  and postconditions requires considerable and careful manual analysis
  with the previous approaches. To extend the analysis to the full
  sample, we use the elastic net approach~\cite{Zou2005}, which combines
  variable selection, regularization and delivering interpretable
  models into one non-parametric, fully automatic approach that is, in
  particular, insensitive to data imperfections such as co-linearity.  The
  analysis widens our results to the full sample of 25 projects, with
  no change in the general statement that \concept has limited to no
  influence on key software qualities.
\end{compactitem}

We judiciously base our analysis on various forms of regression
modeling for two reasons: Firstly, results from any such models are
well \emph{interpretable}, a feature that is not shared by
many of the more recent machine learning techniques, especially as our main
goal is not to make predictions about data by learning from examples,
but to gain an understanding of the measured data sets. Secondly, many
previous discussions of socio-technical issues are based on regression
modelling, and our results can be better put in context and compared
with such efforts by using similar techniques.

However, we would like to point out that the amount of data considered
in our study typically exceeds what previous studies have analysed (as far
as objectively measurable quantities such as number of
developers, amount of source code artefacts, etc.\ are concerned).
In addition, we consistently use time resolution to ensure that changes over
time can be appropriately modelled. This causes a multi-fold increase in the
number of models that need to be computed, and, in particular,
verified and analysed. We need to bridge the gap between using models
that, of course, appropriately represent any insights contained in the
data, but that can also be presented and discussed without overburdening
readers with countless graphs and tables. Furthermore, to exercise
sensitivity analysis~\cite{Shi2005}, we employ a step-wise,
two-fold approach: (1) analysing a selected number of projects in more
depth to establish a baseline understanding of what the data have to
tell, with a focus on establishing model correctness, and then (2)
analysing the complete data set with simpler models, but in more
analysis combinations, to base insights on a broader basis, and to
ascertain that no essential contributing factors have been missed.

Whenever we present the results of a statistical modelling technique
on a subset of the data, we provide the same results and graphs for
all other projects in the sample on the accompanying website
\href{https://cdn.lfdr.de/stmc}{https://cdn.lfdr.de/stmc}, resulting
in hundreds of graphs that can not reasonably be presented and
discussed otherwise. However, great care has been taken to ensure that
the results obtained from subsets can be generalised to the full
sample set by manually iterating over all graphs for each statement
made in the article, and ascertaining that there are no substantial
structural deviations.

\begin{table}[htbp]
  \caption{Covariate magnitude overview for 
    \HBase. Log-transformed quantities are suffixed by ``[l]''. Data
    for the complete set of subject projects are available in the
    online supplement.}\label{tab:covariate_magnitude}
\centering\libertineTabular\rowcolors{2}{gray!10}{white}
  \begin{tabular}{lS[table-format=1.1]S[table-format=1.1]
    S[table-format=1.1]S[table-format=3.0]S[table-format=2.1]}
    \toprule
    {Covariate} & {Min} & {Avg} & {Med} & {Max} & {SD}\\
    \midrule
    \csname @input\endcsname {tab/hbase_overview.tex}
    \bottomrule
  \end{tabular}
\end{table}

To interpret the result of regression models in what follows, in
particular, regarding the \emph{relevance} of individual covariates by
their contribution to the regressor, the magnitude of the covariates
is important. Table~\ref{tab:covariate_magnitude} shows an overview
for the complete dataset of project \HBase.  Values vary for other
projects and depending on the temporal range, but the shown numbers
provide a guideline for ``typical'' values.

\subsection{Multivariate Linear Regression}\label{sec:multivariate_regression}
We focus our attention first on building and interpreting multivariate
linear regression models, as usual in a time-resolved manner. We are
interested in the influence of \concept on key software quality
indicators. However, it is well established
that factors such as lines of code (LoC), number of developers, etc.\
greatly influence the outcomes of interest. This raises the question
to what quantitative degree \measure influences the outcome, and
regression models allow us to infer the influence of \measure while
controlling other factors.

Linear regression models are perceptively easy to specify and compute,
but any conclusions drawn from the results strongly depend on the
correctness and validity of the model specification. One particularly
important precondition is that the amount of collinearity between
predictors is capped. We have used standard techniques of linear
regression models to ascertain this property, as we discuss in more
detail in the Appendix on page~\pageref{sec:model_preconditions}. The resulting
selection of predictors include, beside the various motif-related
quantities that we define in this paper, typical standard
software engineering measures, such as lines of code, traditional
complexity metrics, such as maximal nesting, and socio-technical key
indicators, such as number of developers.

\begin{figure}[htbp]
   \incfig{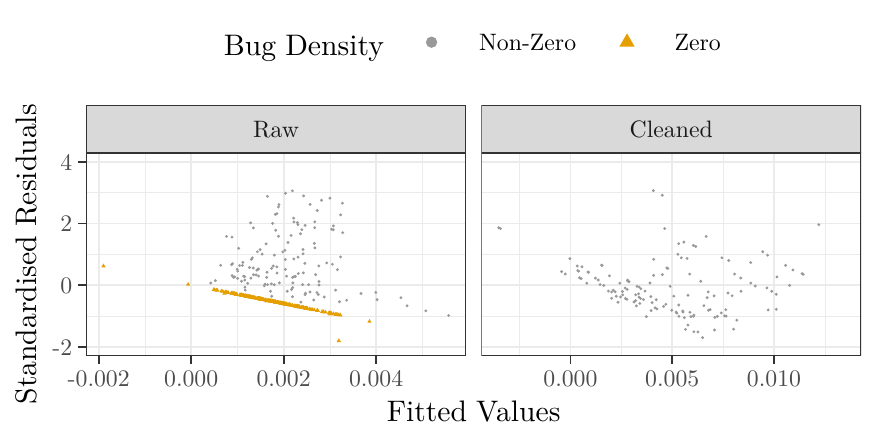}\\[-0.2em]
   \incfig{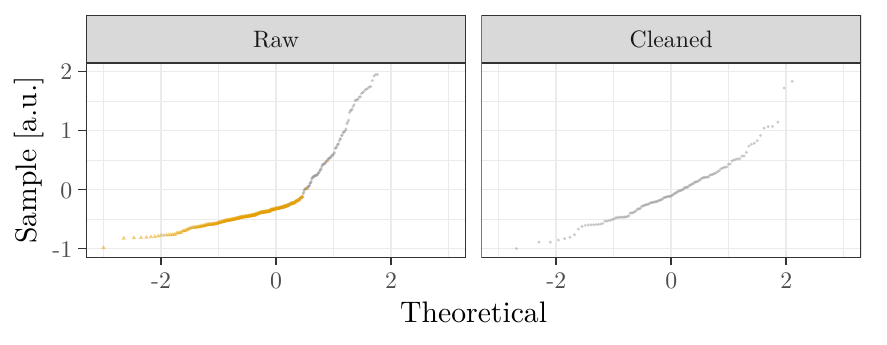}\vspace*{-1.5em}
   \caption{Model diagnostics for the linear model regressing for bug
     density, chosen for an exemplary range (15) of project \Cassandra.  The
     diagnostic plots explore how well the assumptions for
     uncorrelated and normally distributed residuals are satisfied
     (see the text for further details).}\label{fig:bug_density_residuals}
 \end{figure}

 Let us start our discussion by investigating the relation between the
 predictors and bug density, that is, the number of bugs per artefact
 size. Bug density is a metric quantity, and we use a standard
 multivariate linear regression model (see, among dozens or other good
 textbooks, Ref.~\cite{Fahrmeir2013}), given (in matrix notation)
 by
\begin{align}
  \vec{y} &= \mathbf{X}\vec{\beta} + \vec{\epsilon}\label{eq:bug_density_lm}\\
  \vec{y} &= \mathbf{X}\vec{\beta} + \mathbf{Z}\vec{u} +
             \vec{\epsilon}\label{eq:bug_density_lmm}
\end{align}
where each entry of \(\vec{y}\) represents one observation of the
regressand (in more explicit notation, the model can of course be
   written as
   \(y_{i} = \beta_{0}1 + \beta_{1}x_{i,1} + \beta_{2}x_{i,2}+\cdots +
   \beta_{k}x_{i,k} + \epsilon_{i} =
   \beta_{0}+\sum_{i}\beta_{i}x_{i,k}\) for each individual
   observation \(i\) of regressand \(y\) and regression coefficients
   \(\beta_{j}\).).  The associated observations of each regressor (as
determined by the collinearity analysis in
Figure~\ref{fig:correlations_overview} in the appendix) are given as matrix rows in
\(\mathbf{X}\), while \(\vec{\beta}\) collects the regression
coefficients including an intercept term \(\beta_{0}\). Finally,
\(\vec{\epsilon}\) contains the residual values unexplained by the
model, that is, the deviation between measured values and model
result. It holds that
\(\vec{y}\sim N(\Expect{\vec{y}}, \mathbbm{1}\sigma^{2})\).

Discrete regressor quantities with a sufficiently large number of
counts are subjected to the usual variance-stabilising log
transformation, essentially following prior work and common statistical
practise. We indicate log transformations of a given regressor with a
label suffix ``[l]''.

Consider Figure~\ref{fig:bug_density_regression}, which shows the
distribution of regression coefficients for the model covariates, and
the corresponding \(p\) values (see page~\ref{sec:model_correctness}
in the appendix for a discussion on how we ascertain model
correctness; Table~\ref{tab:covariate_magnitude} gives an overview
about typical values that appear in our models as measured for project
\HBase during a three-month long development cycle).  While we include
a (statistically significant) intercept term in all calculations, we
omit it in any graphs of the resulting data, because its magnitude
substantially dominates other regressors.  Besides, the average value
of the regressand at zero contributions of the regressors is not of
interest for our study.  Regardless of what one defines as
``significant'' (we do not interpret significance values for
individual covariates as an indicator of relevance, nor do we mandate
any prescribed ``significance thresholds'' for individual covariates),
we can for \emph{any} of the covariates find an interval where it is
significant (given that \(p\) values are normally distributed when the
null hypothesis of coefficient insignificance is
valid~\cite{Fahrmeir2013}, this is an expected observation).  However,
there are only \emph{two} covariates that consistently show small
values, namely, the number of lines of code, and the number of
developers, both of which are well-known influence factors for quality
properties~\cite{Palomba2018,PalombaBPOPL15}.

Most importantly, none of the socio-technical measures we inspect has
a substantial contribution to bug density, and the contributions of
the quantity is fairly symmetrically centered around zero, which means
that they can have a positive \emph{or} negative influence on bug
density, depending on the analysis interval. Irrespective
of magnitude or statistical significance of the regressors, this means
that increasing the value of \measure by, for instance, establishing a
larger number of positive motifs by changes in development and
coordination processes, can lead to \emph{either} better or worse bug
density while all other influence factors that we consider in our model
are kept constant. This indicates that the measures, and with them the
underlying forms of socio-technical congruence, are not an
optimisation goal worthwhile pursuing.

In addition to the time-resolved multivariate regression model, we
also compute a \emph{mixed} linear model as defined
in~\eref{eq:bug_density_lmm}: Instead of inferring \emph{different}
models for each temporal range, we compute one global \emph{linear
  mixed model} (LMM), but consider the different ranges as an
additional \emph{random} parameter to the model---essentially, this
turns the evaluation of the data sets into a longitudinal
study. This addresses the question of whether the project
  exhibits different characteristics depending on the analysed
  temporal range, respectively of \concept, varies over
  time. Contrariwise, we could assume that a project behaves in
  essentially the same way in all intervals, save for differences
  caused by random, unobserved factors that vary between ranges.
  
  Consider again the results in
  Figure~\ref{fig:bug_density_regression}: red crosses embedded in the
  graphs show the resulting coefficients for the covariates in this
  model compared to the distributions obtained by the standard linear
  model. Most estimates are substantially different from the median
  values of Model~\eref{eq:bug_density_lm}, and many are below the
  first or above the third quantile. This clearly indicates that the
  global, time agnostic model of~\eref{eq:bug_density_lmm} differs
  considerably from time-dependent models of~\eref{eq:bug_density_lm},
  highlighting the need for a fine-grained time-resolved analysis.
  Most importantly, as before: \concept has no appreciable influence
  on bug density, regardless of the measure chosen.

One common, yet debatable measure to judge model quality is the
adjusted \(R^{2}\) value, which indicates what fraction of the variation
in the data is explained by a given model (we use the approach
discussed in Ref.~\cite{Zhang2017} to resolve difficulties with
computing \(R^{2}\) for the more advanced models later on, and to
ensure that computation is based on the same conceptual framework for
all regression models employed in this article). As
Figure~\ref{fig:gam_r_squared} shows, a typical value for \(R^{2}\)
for the linear and generalised linear models is around 0.6, with some
variation among projects and over time.  Regressing for bug density
delivers slightly higher values than for churn, and overall, the
observed \(R^{2}\) values are satisfactory.


We move on to analysing the influence of \measure covariates and more
traditional software engineering metrics for Churn. In contrast to the
previously discussed regressor bug density, Churn is a \emph{count
  quantity} that cannot be assumed to stem from a normal distribution,
and thus necessitates to apply a generalised linear
model~\cite{McCullagh1989,Zeileis2008} that relaxes assumptions on the
response: For one, the expected value may depend on a smooth monotonic
function of the predictors (link function), and the distribution of
the regressor must not be normal, but can stem from an exponential
family distribution,\footnote{Specific members of the family
  \(f_{\Theta}(y) = \operatorname{exp}[(y\Theta - b(\Theta))/a(\psi)
  + c(y, \psi)]\) with scale parameter \(\psi\) and another
  parameter \(\Theta\) include, depending on how values \(a\),
  \(b\) and \(c\) are chosen, the Gaussian, Poisson and Quasi-Poission
  distributions that are relevant for our analysis.} resulting in the
basic structure
\begin{align}
  g(\vec{y}) &= \mathbf{X}\vec{\beta} + \vec{\epsilon}\label{eq:glm},\\
  g(\vec{y}) &= \mathbf{X}\vec{\beta} + \mathbf{Z}\vec{u} +
             \vec{\epsilon}\label{eq:glm_mixed}
\end{align}
where~\eref{eq:glm} represents the generalised linear model (GLM), and
\eref{eq:glm_mixed} is a mixed model extension that allows for including
random effects, as for the linear model of~\eref{eq:bug_density_lmm}. Besides
introducing the link function \(g\) that connects the value obtained
from the predictors by the model with the expected value of the
regressand, the most important change for our purposes is that
\(\Expect{\vec{y}}\sim f_{\Theta}(\vec{y})\), that is, the expected
value of the regressor is determined by the aforementioned
distribution \(f\) controlled by one or two parameters.

A common choice for count data is to use a Poisson model with a
logarithm as link function. The Poisson distribution requires the
expected value of the regressor to be identical to the variance of the
regressor, which we experimentally established to \emph{not} hold for
our data that exhibit overdispersion~\cite{Fahrmeir2013}. Consequently, we
use a quasi-Poisson regression instead, which includes another parameter
(estimated from the data) to model the variance, but is otherwise
mostly identical to a Poisson count regression with log
link. Coefficient estimates will be identical for both
  modelling approaches, but the quasi-Poisson approach leads to an
  adjustment of the inference process for over-dispersed data. In
  fact, a standard Poisson regression on our data leads to significant
  contributions of \emph{all} covariates.

\begin{figure*}[htbp]
  \incfig{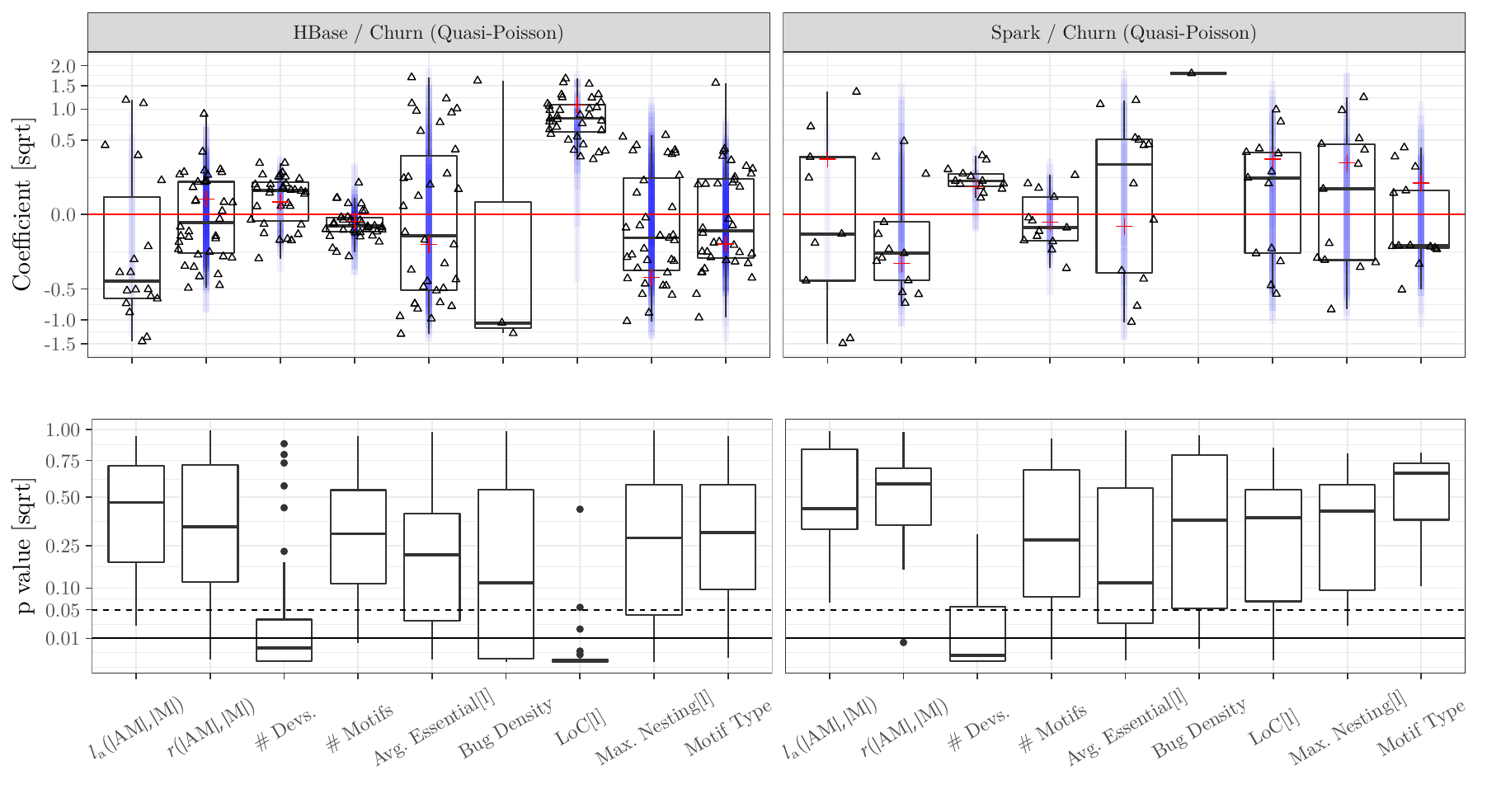}\vspace*{-2em}
  \caption{Results of a time-resolved quasi-poisson logistic
    regression for churn (the figure is restricted to projects
    \HBase and \Spark, but results for other projects --
    as shown in the online supplement -- exhibit very similar
    characteristics).  Boxplots show the distribution of the
    regression coefficients obtained for all time intervals;
    triangles represent the individual coefficient values to provide an
    impression on the actual amount of data.\newline
    The embedded red crosses shows the single coefficient obtained for
    each covariate by a mixed linear models that considers the
    analysis range as an independent stochastic variable. Confidence
    intervals at the 95\% significance level have been computed for
    each time interval; the ranges are shown as partly opaque blue,
    wide solid vertical lines. More intense/darker colouring therefore
    represents regions of an increasing number of overlapping
    confidence intervals. Plots for the complete set of subject
    projects are available in the online
    supplement}\label{fig:logistic_regression}
\end{figure*}

\begin{figure}[htbp]
  \incfig{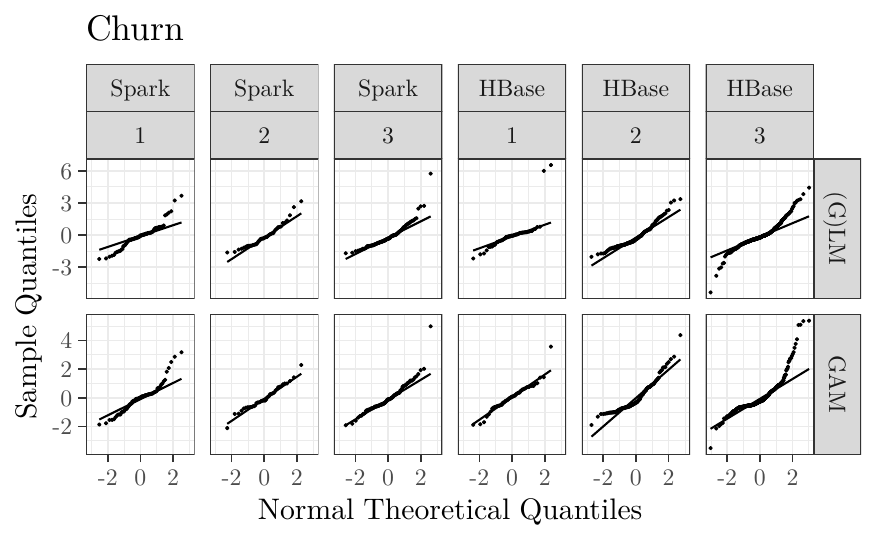}\vspace*{-1.5em}
  \caption{Residual distribution for the quasi-Poisson churn model in
    a quantile-quantile plot. The interpretation of the display is
    identical to Figure~\ref{fig:bug_density_residuals}.
    Deviations from the required normal distribution at the outer
    lobes are present, and indicate the need for improvement.}%
  \label{fig:churn_residuals}
\end{figure}

The results for regressing for churn are shown in
Figure~\ref{fig:churn_residuals}, and paint a similar picture to the
previous regression---most importantly, \concept also has no
influence on Churn. Diagnostic plots as shown in
Figure~\ref{fig:churn_residuals} illustrate that the residual distributions
are (with exceptions for some temporal ranges) satisfactory, and are
likewise with respect to the correlation structure (not shown). Any
inferences drawn from the model are therefore reliable.

\subsection{Generalised Additive Models}
The described model deficiencies, in particular, with respect to the
bug density models, can be eliminated by further lifting
the remaining linearity assumptions of the linear and generalised
linear models, by employing generalised additive
models (GAMs))~\cite{Hastie1990}, formulated as
\begin{equation}
  g(\vec{y}) = \mathbf{X}\vec{\beta} + \sum_{i=1}^{k} \vec{f}_{i}(\vec{x_{i}}) +
                \vec{\epsilon},\label{eq:gam}
\end{equation}
which makes it possible (besides inheriting the properties of the
generalised model including the use of a link function to connect the
expected value of the regressand with the value delivered by the model
from the predictors, and the ability to work with responses that
follow a distribution from the exponential family) that
contributions---from a subset of the covariates---can be modelled by
smooth functions \(f_{i}\) of the covariates. Most importantly, the
optimal ``degree'' of smoothness is non-parametrically determined by
the fitting algorithm, and does not mandate any a-priori functional
choice. While non-parametric transformations necessarily lead to some
reduction in model interpretability, they can be used to cross-check
the aptitude of previously chosen parametric transformations of
covariates.  Technically, \(\vec{f}_{i}(\vec{x})\) applies a smooth
function \(f_{i}(x)\) component-wise to each element of the vector
\(\vec{x}_{i}\), which collects all observations of the \(i\)-th
covariate. We use penalized regression splines, estimated by
penalized regression models~\cite{Wood2017}, to solve~\eref{eq:gam}.

\begin{figure*}[htbp]
  \incfig{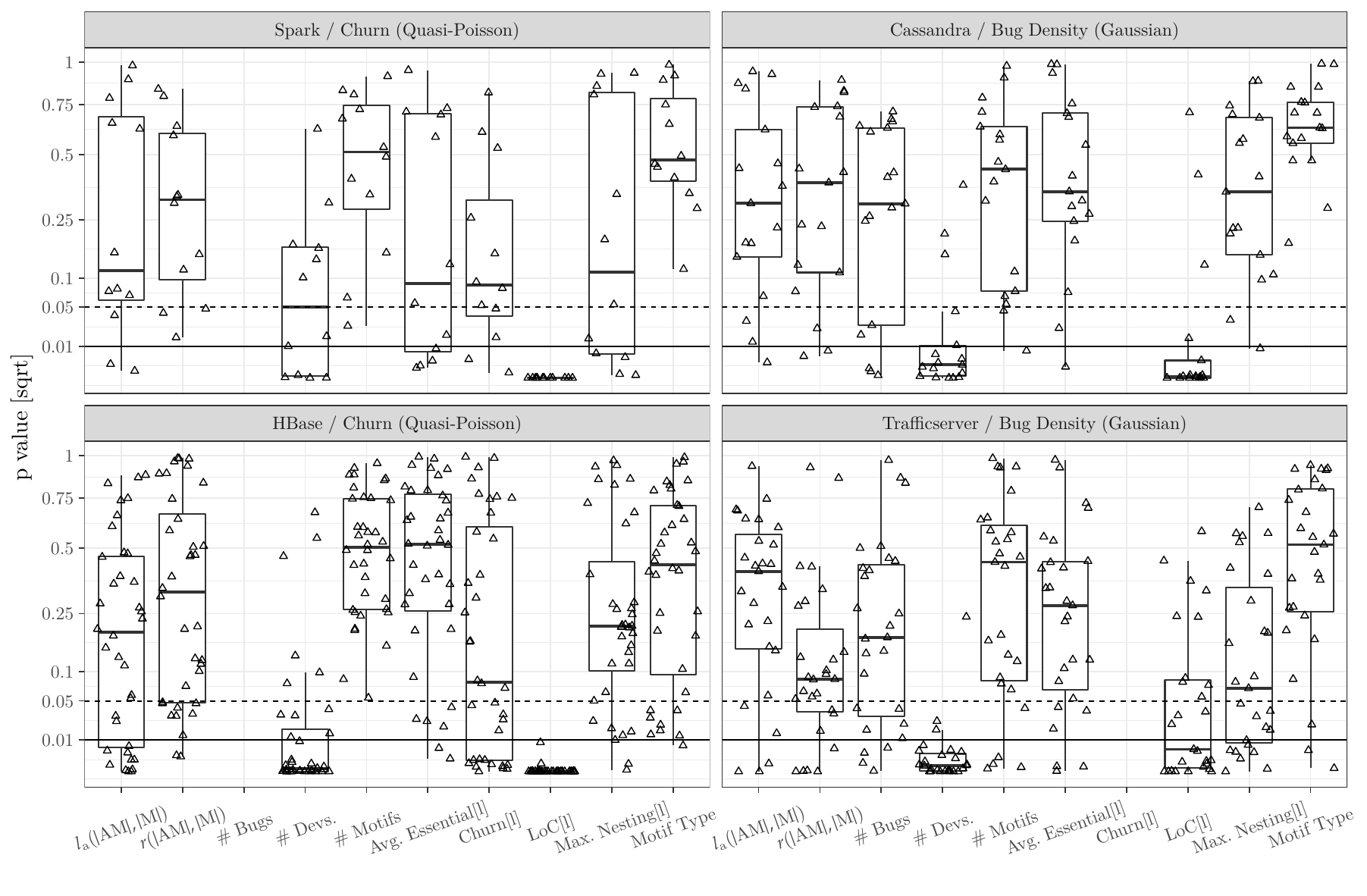}\vspace*{-2em}
  
  \caption{Results of a generalised additive model (GAM) in terms of p
    value distributions for the individual regressors with a
    quasi-Poisson link function for \Spark and \HBase (left)
    with churn as dependent variable, and for a standard GAM Gaussian
    model for bug density for \Cassandra and \TrafficServer
    (right). Smoothing is allowed for all independent variables.
    Triangles embedded in the Box plots represent the actual
    p-values computed for each covariate and each temporal analysis
    range.\newline
    Conventionally employed significance levels of 1\% and 5\%
    are marked by horizontal lines to guide the eye.  Note that
    the \(p\) value scale has been square root transformed to provide a higher
    visual resolution for the range of small values.
  }\label{fig:gam_p_value_distrib_bug_count}
\end{figure*}

\begin{figure}[htbp]
  \incfig{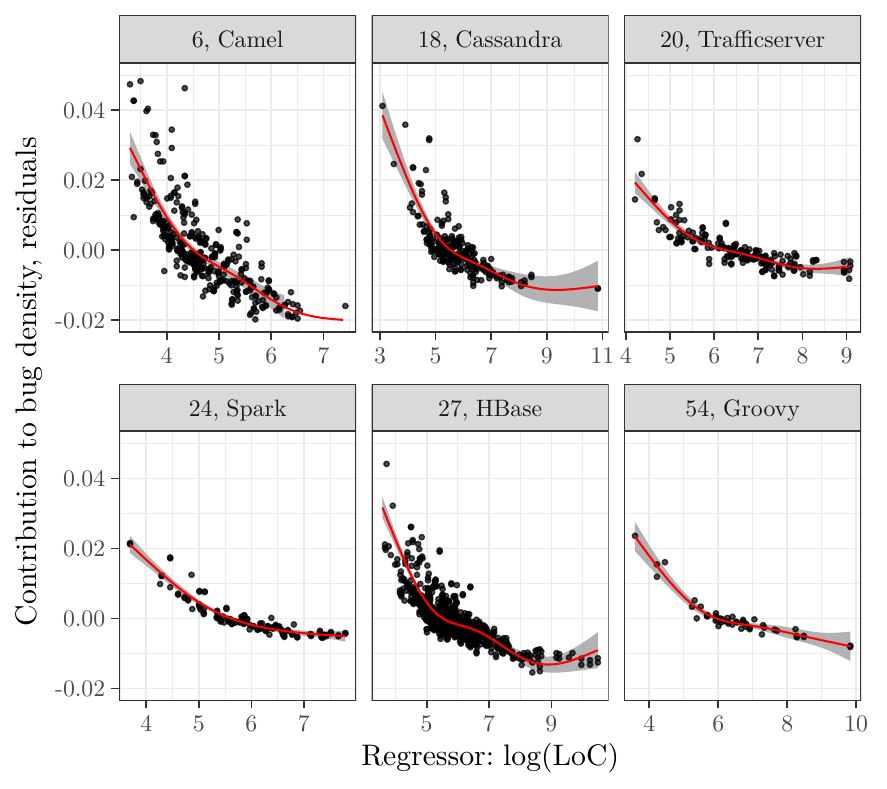}\vspace*{-1em}
  \caption{Illustration of non-linear transformations on the
    contribution provided by LoC[l] by GAM (the shaded areas
    represents the 95\% confidence interval). The transformation is
    structurally similar for the shown revision ranges and projects:
    For small amounts of source code, bug density increases for more
    code since the coefficient is positive; however, the \emph{rate}
    at which bug density increases with growing amounts of code
    quickly drops for larger source files. From a certain threshold
    onwards (5, representing \(10^{5}\) LoC owing to the log transform
    of the predictor), the influence of LoC to bug density is getting
    less pronounced, and can even become \emph{negative}. This is
    consistent with observations by other
    authors.
    More importantly,
    though, effect and transformation chosen by the model are not only
    consistent over different time ranges, but also across projects,
    which increases confidence in the consistency of our 
    data.}\label{fig:gam_trafo_illustration}
\end{figure}

\begin{figure}[htbp]
  \incfig{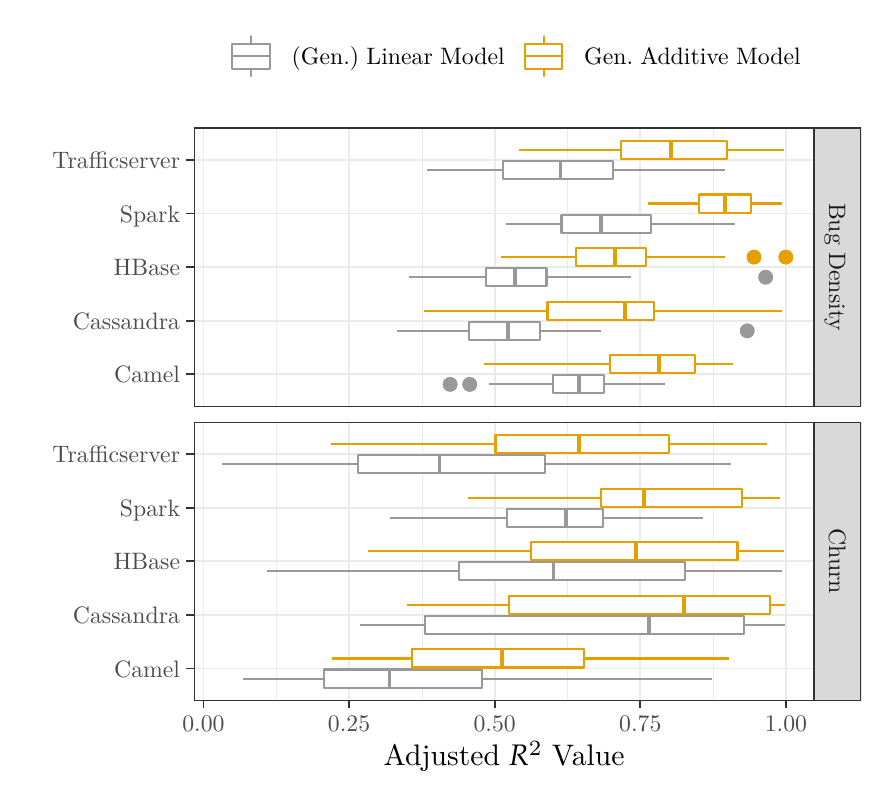}\vspace*{-1.5em}
  \caption{Adjusted \(R^{2}\) values for (generalised) linear and
    generalised additive regression models, computed for bug density
    and churn as regression targets. The box plots show the
    distribution of values for a time-resolved application of the
    models. Notice the cautionary remarks on the limited utility of this
    omnipresent measure in the text.}\label{fig:gam_r_squared}
\end{figure}

The conclusions that can be drawn from the generalised models
(cf.~Fig.~\ref{fig:gam_r_squared} and
\ref{fig:gam_p_value_distrib_bug_count}) are three-fold: Firstly,
introducing non-linearity into the model considerably improves the
\(R^{2}\) measure to values usually well above 0.8, which is more a
testament to the bounded amount of noise in the data than to model
correctness, considering that only low-dimensional non-linear
transformations have been chosen by the model (better values can, in
general, not be expected for processes involving human
participation~\cite{Moksony1999}, so in this sense, our model is sufficiently
complete). However, the value may be helpful for readers to broadly
relate our results to related work that specifies model quality only
in terms of \(R^{2}\). GAMs also offer a clear improvement in terms of
\(R^{2}\) over generalised linear models. While in itself this is
uninteresting (adding covariates, which a non-linear transformation is
essentially bound to do, will always improve \(R^{2}\)), the fact that
the non-linear transformation is essentially identical for all
projects and all temporal ranges, as shown in
Figure~\ref{fig:gam_trafo_illustration} for a subset
(\Spark, \HBase, \Camel, \TrafficServer, \HBase and \Groovy) demonstrates
that the transformation is structurally similar for varying revision
ranges and projects. The non-linear transformations only affect
quantities that anyway dominate the previous analysis, namely LoC and
developer count, which ascertains that the lack of relevance for
\concept measures does not stem from ill-specified or missing data
transformations, but is inherent in the data as such.

The improvement obtained by non-linearly transforming the covariate
beyond the usual variance-stabilising transformation is limited, and
it does not uncover the need to use a substantially different a-priori
data transformation than then ones already employed. As a minor
consequence, we note that the data set could probably also be used for
predictive purposes after symbolically modelling the transformation
that is chosen by the non-parametric
approach. Figure~\ref{fig:gam_trafo_illustration}, as compared to
Figure~\ref{fig:gam_r_squared} strongly hints that missing
explanatory variable data are not an attractive possible cause of
deficiencies of the linear and generalised linear model resulting in
moderate \(R^{2}\) values, since the required non-linearities are not
severe.

To summarise our main findings: \concept does not provide a substantial
explanatory value also for GAM models -- the influence is on par with
known problematic~\cite{Shepperd1988} predictors such as essential and
cyclomatic complexity, and does therefore not seem relevant for practical
software engineering purposes.

We carefully evaluated the quality of the approach's residual
structure using similar tests as before to ensure mode correctness,
but refer the reader to the online supplement for details and
graphical summaries.  As always, full results for other subject
projects are available in the online supplement.

The choice of generalised additive models leads to a much improved
model quality compared to more straight-forward regression approaches
(the automatically chosen nonlinear transformations do, of course,
reduce the predictive power of our models to some extent, but
prediction is not a goal of this work---our study is for one
confirmative, and prediction in the absence of an effect is anyway of
limited value), which is once more underlined by
Figure~\ref{fig:gam_r_squared}, which shows the value distribution of
adjusted \(R^{2}\) values. It is much improved compared to the results in the
previous section. Additionally, it puts our models on par (or even
improves over) the findings in the seminal work of Cataldo, Herbsleb
and coworkers~\cite{CataldoH13,Cataldo2008} with respect to this
quantity and the amount of required regressors. We emphasise once
more the restricted value of \(R^{2}\) in assessing model
quality despite its wide-spread application in the software
engineering literature, and the clear limitations of such comparisons.

\subsection{Model Scale-Up}
The modelling approaches that we have employed so far, and also many
of the approaches used in the literature, use various forms of
regression analysis to establish results that balance two essential
needs of statistical analysis: Firstly, to obtain good explanatory (or
even predictive) performance, and secondly, to provide a parsimonious,
interpretable model that can be discussed (and tested!) using a-priori
scientific knowledge or human understanding. This requires careful
mathematical modelling and, more importantly, comprehensive
diagnostics to ensure correctness of statements derived from the
results.  Drawing conclusions and generalisations from a large number
of samples is a daunting enterprise in that context. We will address
this problem in the following.

The needs of establishing correct and parsimonious models require
selecting \emph{relevant} variables without any a-priori fixing of
(essentially arbitrary) significance values, and the correct handling
of problematic structures in the data, which can lead to ill-specified
models. Consequently, we now focus on effective and robust models, and
judge covariate importance not based on significance (effectively also
removing sample size considerations), but on predictive importance and
model performance, even if our concern is \emph{not} prediction, but
understanding.  In a way, this shifts our ``modelling
culture''~\cite{Breiman2001} from algorithmic modelling towards data
modelling. Eventually, the conclusions that we can draw from employing
the two philosophical approaches will turn out to be identical, which
we feel is a scientifically reassuring result.

Ordinary least squares (OLS) and generalised regression is known to
often perform badly when trying to solve both purposes, description
and prediction~\cite{Zou2005}. Commonly employed automatic methods to
achieve parsimony (like, for instance, best subset selection) have
been shown to be afflicted with numerous issues such as instability
(see, \eg,~\cite{Breiman1996}. Ridge regression~\cite{Hoerl1970}
augments OLS with a penalty on regression coefficients measured by the
\(\text{L}_{2}\) norm\footnote{Recall that
  \(\|\vec{x}\|_{p} \coloneqq
  \left(\sum_{i=1}^{m}|x_{i}|^{p}\right)^{1/p}\) defines the
  \(p\)-norm of an \(m\)-dimensional vector \(\vec{x}\) that reduces
  to the usual Euclidean norm for \(p=2\).}  \(\|\cdot\|_{2}\). It
solves problems with multi-collinearity by evading matrix inversion
singularities, and consequently requires no up-front removal of
similar (or even identical) predictors, supporting the requirement for
a fully automatic analysis process. However, ridge regression always
retains the full set of regressors in the model, thus failing the
parsimony requirement. The latter is achieved by a structurally
similar approach, lasso regression~\cite{Tibshirani1994}, which
penalises regression coefficients based on the \(\text{L}_{1}\) norm
\(\|\cdot\|_{1}\). Lasso regression can perform model shrinking by
automatic selection of covariates, but is afflicted with the issue of
essentially randomly picking one regressor from a group of correlated
regressors, among other less important drawbacks~\cite{Zou2005}.

Eliminating the drawbacks and combining the advantages of both methods
is done by the \emph{elastic net} approach~\cite{Friedman2010}, which
interpolates between ridge and lasso regression. Based on the same
specification as for orthodox OLS given in
\eref{eq:bug_density_lm},
\(\vec{y} = \mathbf{X}\vec{\beta} + \vec{\epsilon}\), the elastic net
determines coefficient estimates \(\vec{\beta}\) by solving the
problem
\begin{equation}
  \hat{\beta} \coloneqq {\underset{\beta}{\operatorname {argmin}}}\ 
  \lambda\left[(1-\alpha) \|\vec{\beta}\|_{2}^{2}/2 +
    \alpha\|\vec{\beta}\|_{1}\right]\label{eq:elastic_net}
\end{equation}
subject to hyper-parameter \(\lambda\in [0,1]\), which controls the
strength of the penalty, and a ``mixture'' parameter
\(\alpha \in [0,1]\) that bridges between the cases of pure ridge
(\(\alpha = 0\)) and pure lasso (\(\alpha = 1\)) regression. Note that
the elastic net allows us to work with responses that stem from
Gaussian or Poisson distributions. It is not possible to employ the
Quasi-Poisson family to account for differing variance and expected
value, which is the case for our data. Since we are anyway unable to
provide confidence intervals---that would be invalidated by this
shortcoming---for elastic net results, and the point-wise estimates
are identical for Poisson and Quasi-Poisson based computation, this is
not a concern for our analysis.

For each time window, we separately determine the best set of values
for \(\lambda\) and \(\alpha\) by using a cross-validation
approach~\cite{Hastie2009}, which delivers the smallest mean error of
the resulting model. We do not consider values of \(\alpha\) outside
of \([0.1, 0.9]\) to avoid any of the drawbacks of the boundary cases
discussed above. This does not introduce an arbitrary choice into our
analysis---the optimisation procedure turns out to always deliver
values in this range anyway. We have installed warning mechanisms that
give note in case a pure ridge or lasso would result as best
performing procedure.

In the following calculations, we scale and center the data so that
they exhibit zero mean value and unit variance, which is a common, yet
not undebated practice in regression analysis to compare the
\emph{relative} influence of coefficients (Ref.~\cite{Baguley2009}
reviews the many arguments for either side) to compare the
relative ``importance'' of regressors, which gives a reasonable means
of gauging the relative importance of predictors~\cite{Agresti2007}: A
change by \emph{one unit} of each regressor (while keeping the other
variables in the regression model constant) corresponds to a change by
\emph{one standard deviation} of the regressand, However, the reader needs to
keep in mind that non-linear transformations are applied to the
predictors \emph{before} scaling and centering. Since most of the
criticisms on the transformation relate to models with non-linear
contributions, interaction effects, or handling of categorical
variables, we refrain from introducing the first two in the following,
and perform separate analyses with respect to the motif type (square
or triangle), which is the only categorical variable in our data set.

One drawback of the elastic net approach (and, partially, the
non-standard inference approach via cyclic coordinate
descent~\cite{Friedman2010}) is that the question of how to provide
confidence intervals, which we have done for all approaches so far, is
currently still subject to considerable scientific
debate~\cite{Kyung2010}. Remedies based on bootstrapping and
resampling usually result in too small quantities~\cite{Goeman2010},
which give a false sense of precision. Consequently, we discontinue
the provision of confidence intervals, contrary to previous sections.
  
\begin{figure*}[htbp]
  \incfig{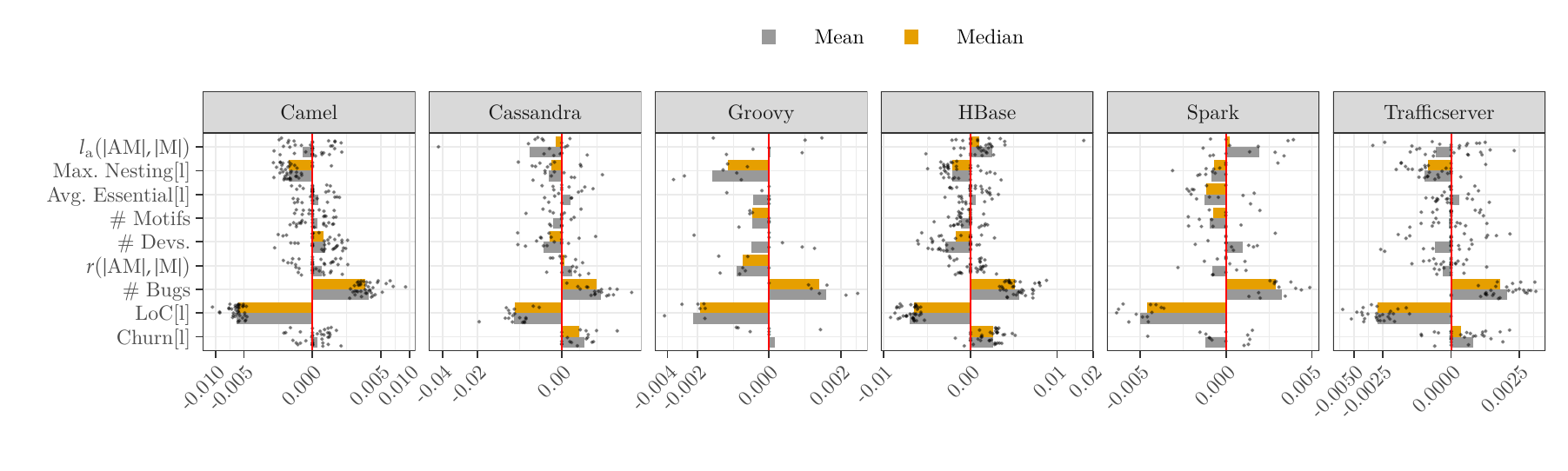}\\[-1em]
  \incfig{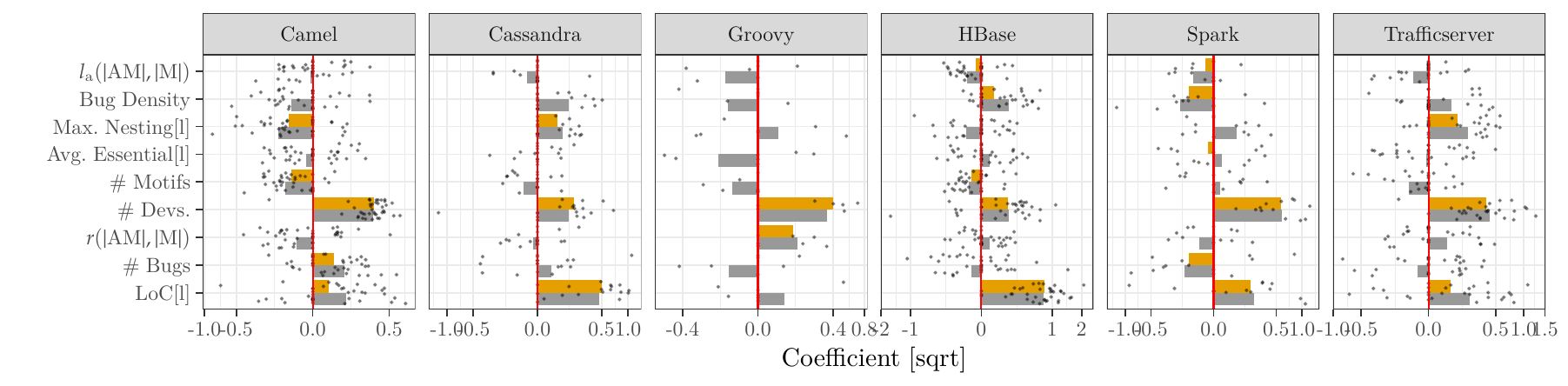}\vspace*{-1em}
  \caption{Coefficient distribution for elastic net regression
    models (top row: bug density, bottom row: churn). Every point
    represents the magnitude of a regression coefficient for one time
    interval (based on scaled and centered input data), and bars
    show the mean/median value over all analysis intervals. A
    two-sided square root transformation is used on the coefficient
    values, and the red vertical line highlights coefficient value
    0. Plots for the complete set of subject projects are available on
    the supplementary website.}\label{fig:lasso_variable_selection}
\end{figure*}

For both regressands---bug density and churn---large values are
undesirable, and both \measure measures \emph{shrink} (in the sense of moving
towards negative values with larger absolute magnitudes) when the
relative number of motifs \emph{increases}, and \emph{grow},
respectively become more positive when the relative number of motifs
increases:
\begin{align}
  \text{\concept } (r < 0) \Leftrightarrow \text{fewer bugs},\\
  \text{Anti-\concept } (r \geq 0) \Leftrightarrow \text{more bugs}.
\end{align}
The same applies to \(l\). This translates into an expected
\emph{positive} relation between the regression coefficients for \(r\)
and \(l\) and bug density, and likewise for \(l\).  The convention
ensures that \emph{positive} regression coefficients indicate a regime
that we denote as STMC regime, and negative
coefficients arise in the opposite, anti-STMC regime.

The results of computing the model for the set of the usual six reference
projects that can still be conveniently visualised are shown in
Figure~\ref{fig:lasso_variable_selection}. The figure is intended
to make statements about two (time-wise) global and local aspects:
Results for each time interval are given by individual data points,
and for each interval, the model delivers \emph{one} value for \emph{each}
covariate. To see whether the behaviour is consistent over time, we
need to aggregate the local results into one global result. We
do this by computing the median of the results over time for each
covariate (since mean and median summaries
of the coefficient distributions agree well, the model does not
deliver results that strongly deviate from the total set of results,
which shows result stability, an important property in automated
analyses). The same covariates that have consistently resulted in small
\(p\) values for the previous approaches are assigned coefficient
magnitudes that differ from zero. The magnitudes for both realisations
of \measure are small and, more importantly, scatter around zero,
which once more underlines that their influence on software quality
can be both good and bad. Consequently, they are inopportune
quantities for the goal of  optimising development processes.

Finally, we would like to draw attention to the consistency of
results across projects (easily seen by comparing the columns of each row of the
graph), which once more underlines that our observations are not
peculiarities of individual projects, but relate to general properties
of software projects.

In summary, the results obtained for the sample set are consistent
with any of the results of the in-depth analysis, which provides
confidence that applying the method to the full project set in the
same analysis setting delivers reliable outcomes, which we feel once
more strengthens our conclusions.

\subsection{Large-Scale Elastic Net Deployment}
None of the previously considered sample projects has shown any
relevant influence of \measure so far. However, this might still be
caused by particular properties of the projects, and do not generalise
to a larger sample. To decisively answer RQ2 (and because
generalisability is at the core of most meaningful scientific
statements), we want to place our insights on a considerable larger
sample space introduced in Table~\ref{tab:subject_projects}. This
requires us to compute the elastic net not only for more projects than
before, but also in three different temporal scenarios, resulting in
a number of combinations that can not reasonably be visualised as in
Figure~\ref{fig:lasso_variable_selection}. Consequently, we compress
the display of results in two more stages.

\begin{figure*}[htbp]
  \incfig{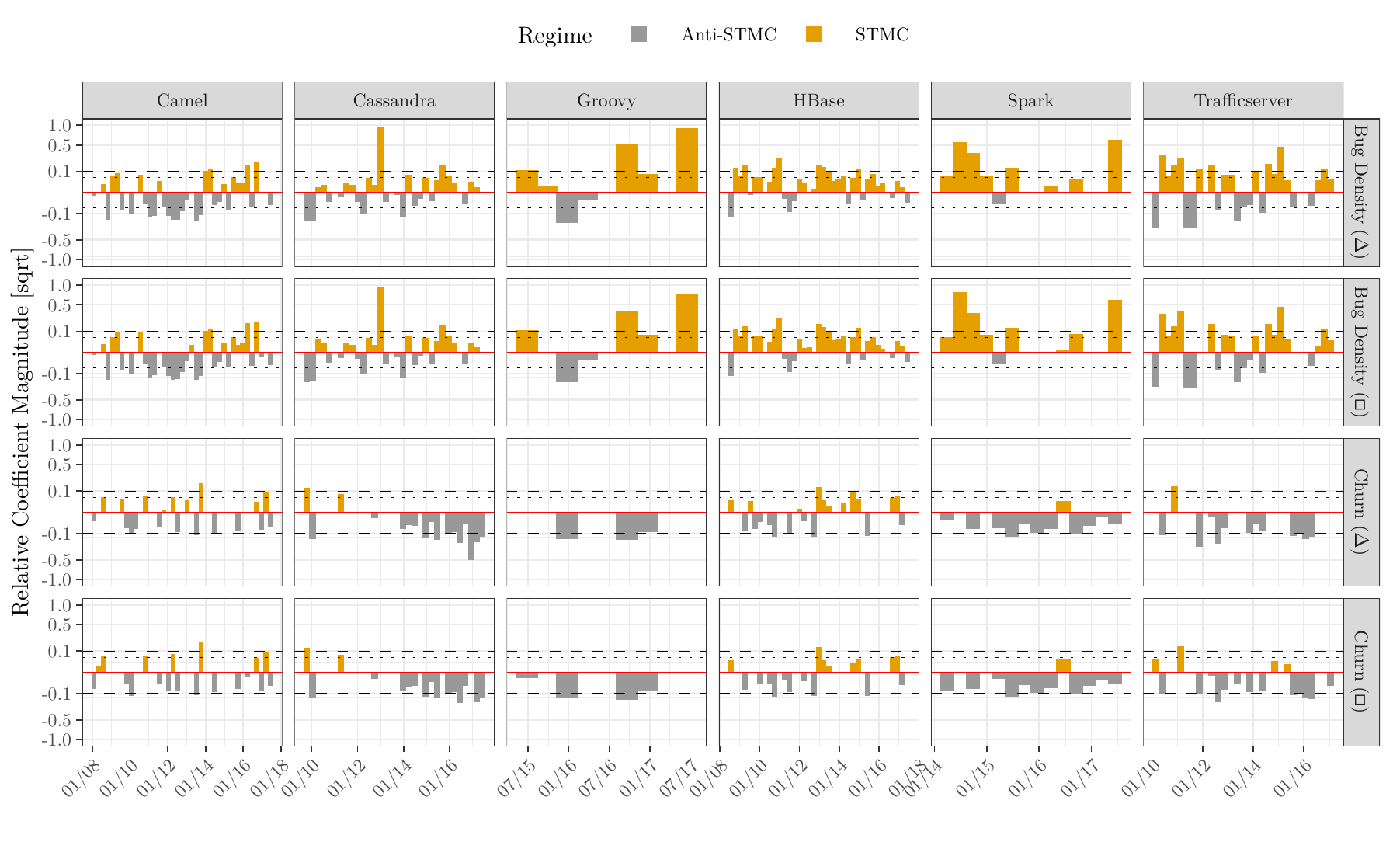}\vspace*{-2em}
  \caption{Isochronous time series relating \concept regressors and
    predicted quantity (bug density of churn) for a subset of
    subject projects based on combined email and Jira
    communication, as well as file-level artefact dependencies obtained by
    co-change relationships---data and graphs for the full set of
    projects are available at the supplementary website. Dotted
    (\(\pm 0.05\)) and dashed (\(\pm 0.1\)) lines indicate corridors
    for 5\% and 10\%, but are only used to guide the eye and not as a
    dichotomic decision boundary. Symbols \(\Delta\) and \(\square\)
    represent triangle and square motif, respectively. Results for all
    subject projects are available on the supplementary website.}\label{fig:corr_ts}
\end{figure*}

\newcommand\varpm{\mathbin{\vcenter{\hbox{%
        \oalign{\hfil$\scriptstyle+$\hfil\cr \noalign{\kern-.5ex}
          $\scriptscriptstyle({-})$\cr}%
      }}}}


Figure~\ref{fig:corr_ts} provides a time series for each of the four
combinations of motif (triangle/square) and regressor (bug
density/churn) for one particular analysis combination---email+Jira
communications with co-change dependencies in an isochronous temporal
relation (we have performed the same calculations with reduced
communication data sets, which did not result in substantial
differences to the shown results, indicating that our data are
complete in this aspect). For each data point, we compute the sum of
all \emph{absolute} values of the regression coefficients, and then we
infer the \emph{relative} magnitude that the coefficient for
\(l(\AMotif, \Motif)\) contributes. This means that each data point
represents \(\beta_{k}/\sum_{i}|\beta_{i}|\), where \(k\) is the index
of the coefficient for \(l\). The magnitude of the quantity measures
the extent of influence of \measure on the regressor, and the sign
shows if the influence is beneficial (+) or adverse (-).

The observed magnitudes rarely exceed 10\%, and often remain below
5\% (time points with no visible contributions can either indicate a
value that was too small to plot, despite the symmetric square root
transformation that magnifies smaller values, or that the elastic net
assigned a coefficient of zero). While positive, beneficial
contributions occur slightly more often than negative ones (with the
exception of bug density and the triangle motif for project \HBase,
where positive contributions that also consistently dominate), there
is an overall balance between positive and negative effects without
any recognisable temporal patterns. This indicates that the presence
of \concept has not only has limited influence, but can also,
essentially randomly, at one time cause benefits, and at another time
disadvantages.

The correlation time series also illustrates another aspect: Despite
the well-known substantial correlation between bugs and
churn~\cite{Zimmermann:2008}, the time series can differ considerably
when the influence of the \concept measure is compared for the same
motif, but with different regressands. This is nicely visible for the
projects \Spark and the triangle motif: The consistently beneficial
contributions with respect to bug density turn into a consistently
negative contribution for churn. Comparing the results for identical
regressands but different motif types, shows more consistent, but also
clearly different behaviour. This underlines that both motifs and
quality measures capture different aspects of the projects, and do not
just re-iterate the same observations.

\begin{figure*}[htbp]
  \incfig{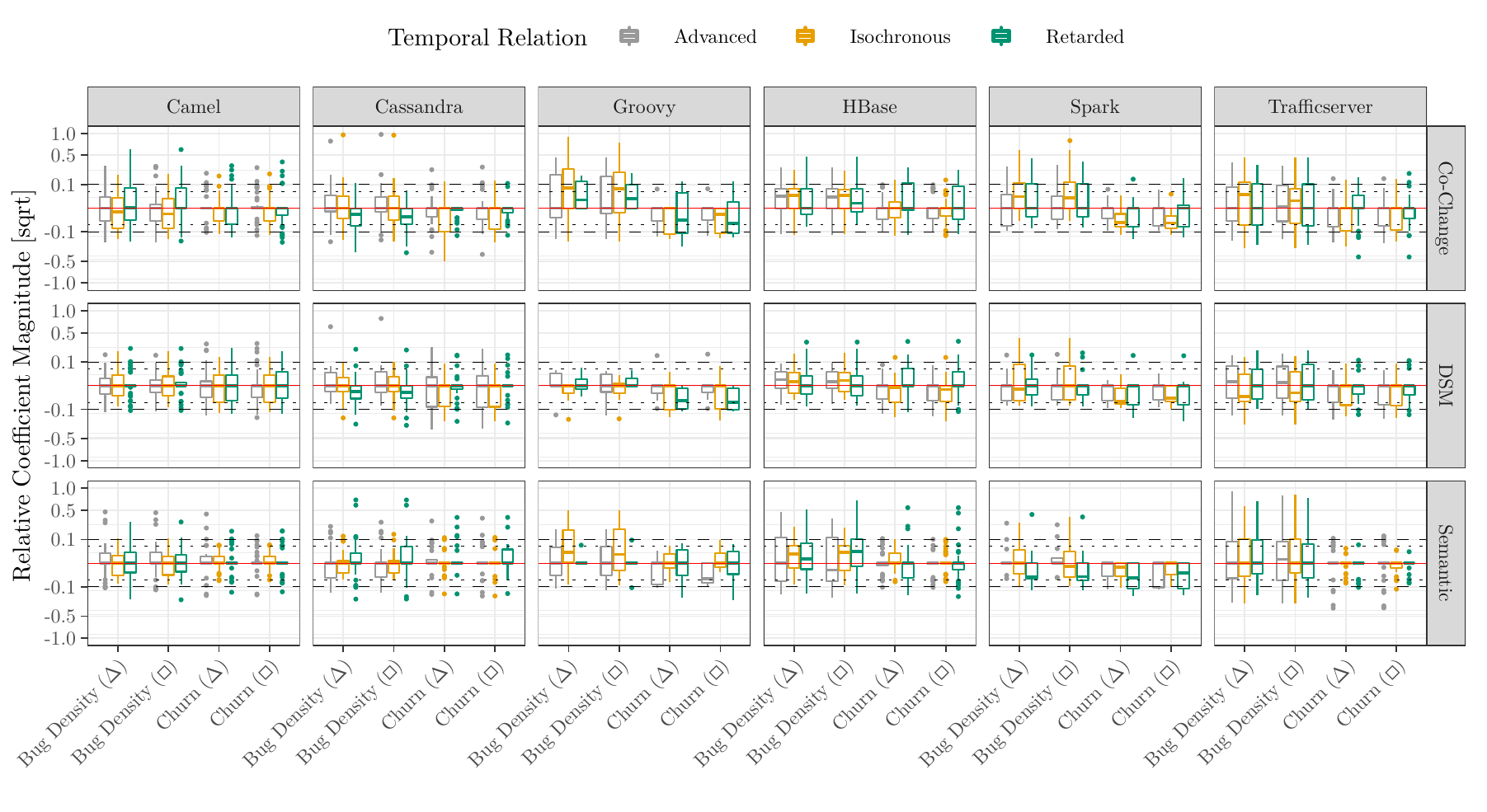}\vspace*{-2em}
  \caption{Relative coefficient magnitude (influence of motif LoC norm
    diff \(l_{\text{a}}(\AMotif, \Motif)\)) time series summaries for
    a typical subset of subject projects for correlations based on
    email and Jira communication data for different temporal
    relationships (isochronous, advanced and retarded). Horizontal
    lines are as in Figure~\ref{fig:corr_ts}}\label{fig:tt_overview}
\end{figure*}

\begin{figure*}[htbp]
  \incfig{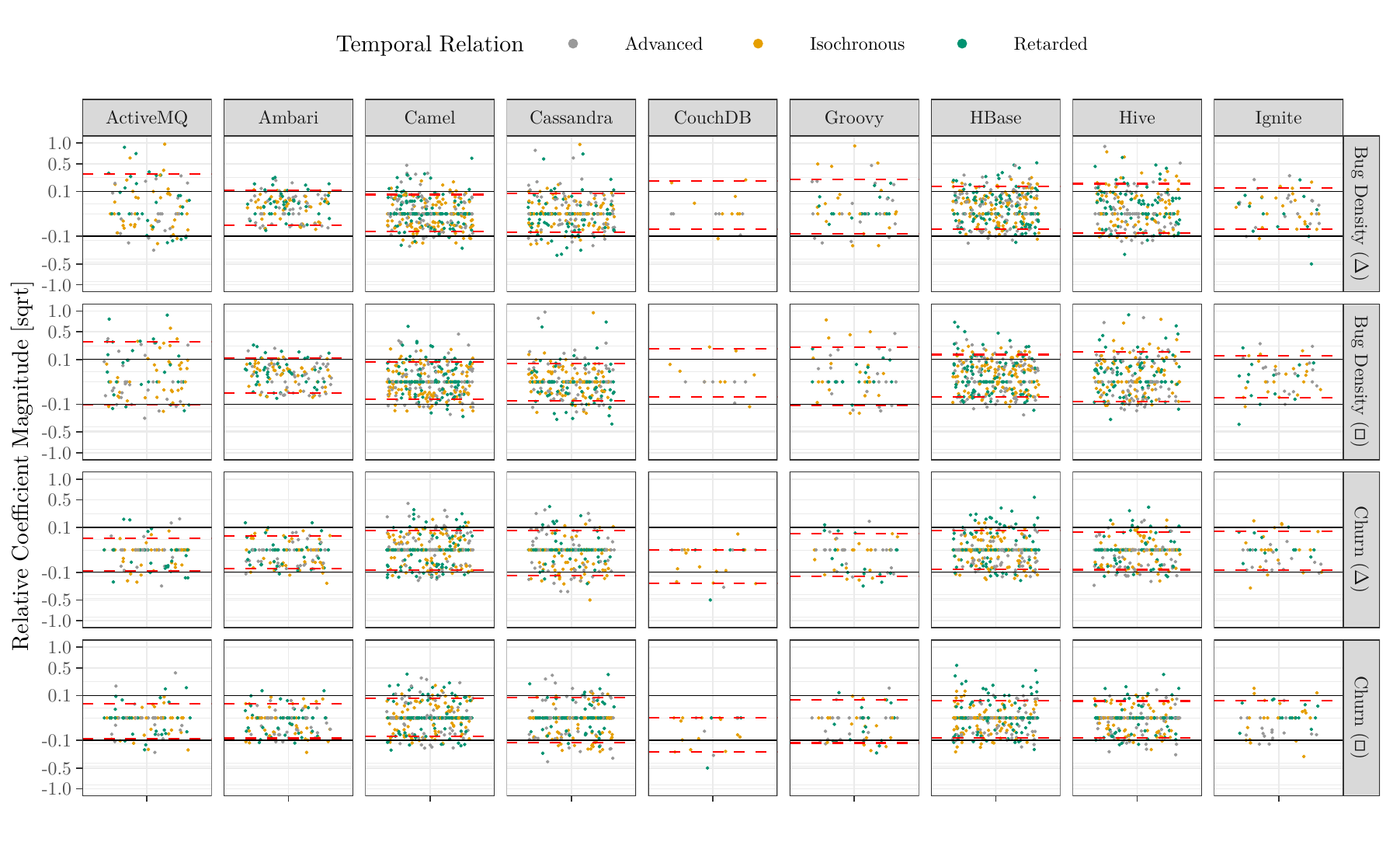}\\\vspace*{-3em}
  \incfig{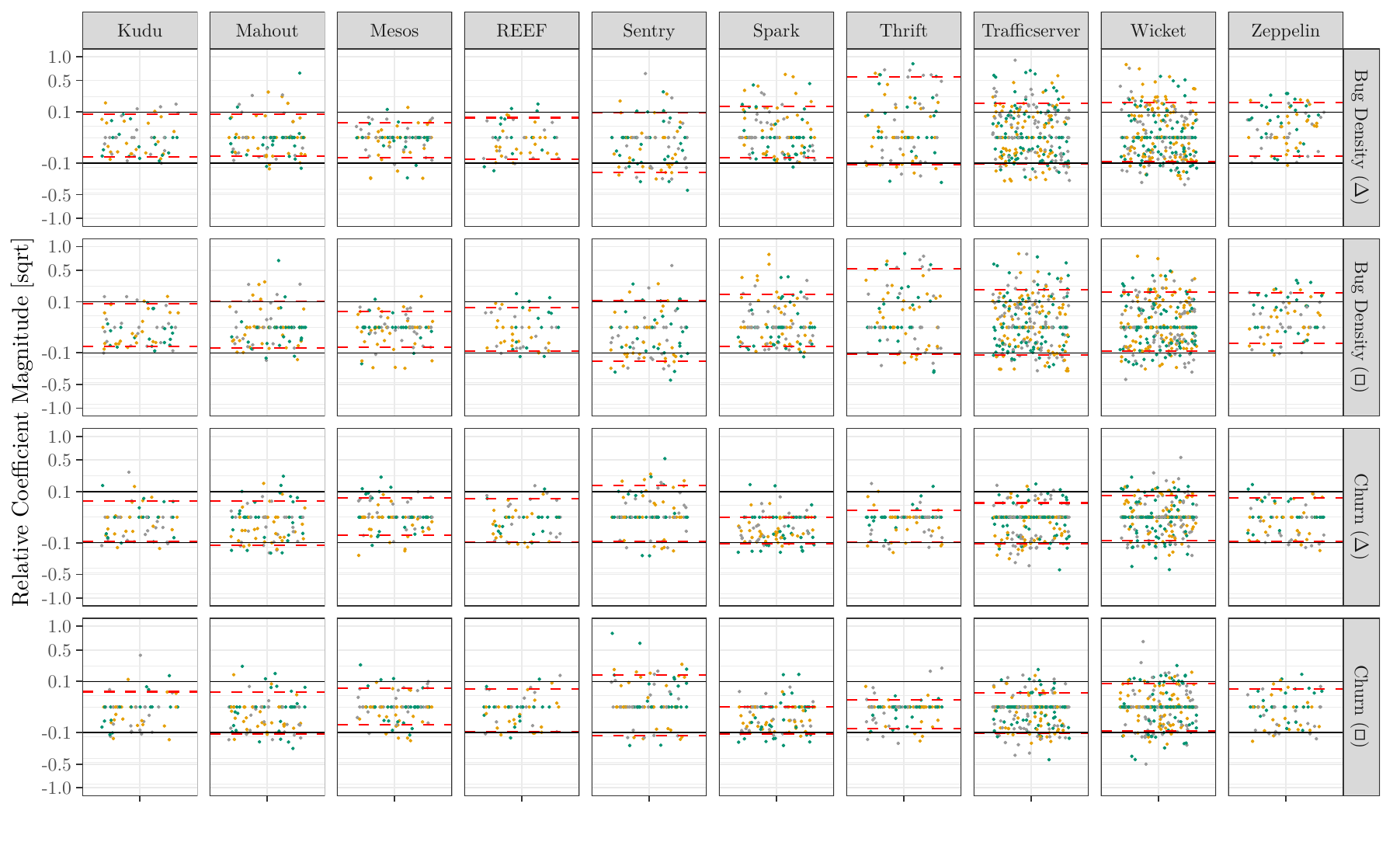}\\
  \caption{Summary of relative coefficient magnitudes for
    \(l_{\text{a}}(\AMotif, \Motif)\), displayed for the complete
    subject sample set, all temporal directions (advanced, isochronous,
    retarded), quality measures (bug density, churn) and coupling
    mechanisms (co-change, semantic coupling, DSM). Observe the square
    root transformation of the magnitudes on the y scale to enlarge
    the densely populated region centered around zero. The horizontal
    spread per project is caused by jittering the data points to
    ensure visibility of the complete data set. The solid horizontal
    lines indicate the corridor for \(\pm\) 10\%; dashed red lines
    show the 10 and 90 percent quartile of the
    data.}\label{fig:tt_all}
\end{figure*}

Results for the most comprehensive set of communication data
(email+Jira) and all three dependency mechanisms (co-change, DSM, and
semantic dependency) are shown in Figure~\ref{fig:tt_overview} for
a subset of sample projects. One boxplot in
Figure~\ref{fig:tt_overview} summarises the content of one single
panel in Figure~\ref{fig:corr_ts}. It is also immediately obvious that
no substantial differences arise from the use of different coupling
mechanisms, which means that our key observations are valid regardless
of the exact construction details of the underlying socio-technical
network.

The corresponding results for the full set of sample projects are
given in Figure~\ref{fig:tt_all} (it suffices to consider the set of
isochronous data points for now): Each data points represents the
relative influence of \(l_{\text{a}}(\AMotif, \Motif)\) on the
regressand; dashed lines indicate the 10\% and 90\% quantiles,
which means that the dominant fraction of all observed values is
confined between these two lines. By comparing these lines with the
solid horizontal lines that indicate boundaries for \(\pm\)10\%, it becomes
clear that the influence of this socio-technical measure on
quality is very weak. Even in the cases where consequences of \concept
hold, they do not have much influence on software quality as measured
by bug density. The results for churn are essentially identical, and
likewise for other measures of socio-technical congruence. Space
restrictions require us to refer to the online supplement instead of
presenting the graphs here.

\subsection{Answering Research Question 2}
Considering the complete set of analyses performed on the data, we
can summarise that \concept does not have any notable influence on
the robust and widely deployed quality indicators of bugs and churn. 
We must therefore answer RQ2 \emph{negatively}.


\section{Research Question 3---Temporal Analysis}
The previous discussion was concerned with relating different
co-variables from the \emph{same} time window. This has shown that
\emph{no} meaningful relationship between \concept and software
quality indicators exists. This suggests the interpretation that
the presence of \concept delivers no \emph{short-term} effects on
software quality. However, this does not exclude the possibility that
\concept could lead to significant effects only after a time
delay. More generally, \concept in one time window could be
correlated with observable effects in a different time window. 

To determine if there are any such time-shifted relations, we
introduce \emph{advanced} elastic net models that consider quality data in
time window \(n\) and motifs in the following time window \(n+1\), as
well as \emph{retarded} models that relate quality data in time
window \(n\) with motifs in previous time window \(n-1\) (we refer to
our prior analyzed correlations, that connect co-variables from the
same time window, as \emph{isochronous} in the
following).  Intuitively, the presence of retarded relations
suggests that a positive amount of \measure leads to improvements in
project quality at a later time.  Strong advanced relations, on the
other hand, could be interpreted as a situation where the presence of
a bug, a complicated piece of code, or other undesirable circumstances
leads to more communication at a later time because discussion among
developers is required to find resolutions.

Results for advanced and retarded elastic net model calculations are
given in Figure~\vref{fig:tt_all}. As in the isochronous case, it can
be seen that both, \concept and Anti-\concept regimes are present over
time, but with only moderate to negligible relative influence of the
socio-technical regressors. Recall that each data point shows the
relative contribution of \(l_{\text{a}}(\AMotif, \Motif)\) to various
models; it is visually obvious that almost no data points exceed a
relative contribution of 10\%, and---as the red lines that denote the
10 and 90 percent quartiles show---the relative influence is usually
even less than this amount. All these statements hold regardless of
temporal direction. The results therefore effectively confirm the
outcome for the isochronous models as discussed in RQ2.

One minor, but notable observation is that the relative magnitude of
socio-technical indicators is slightly stronger for isochronous than
advanced and retarded correlations, which intuitively hints at the
interpretation that effects of communication and cooperation on
software quality do not manifest over arbitrary time distances, but
materialise in temporal proximity. This shall be explored in further
work.

\subsection{Answering Research Question 3}
In summary, we see that the maximal deviations for non-isochronous
temporal relations are insubstantial and permit us to draw a
conclusion analogous to RQ2: The influence of temporal relationship is
insubstantial, as the observed relations between \measure and software
quality indicators are again small to negligible. \concept is not a
significant and relevant predictor for future software quality, and
past software quality is not related with future \concept.


\section{Verifiability \& Threats to Validity}\label{sec:threats}
Our work, like all research efforts, rests on certain assumptions
that must be taken, and may be affected by factors that are out
of our experimental control. We turn our attention to these
aspects next.

\subsection{Verifiability}
Our methods have been consistently designed for independent external
verification and replication.  All code and analysis infrastructure
(with the exception of the Understand tool, for which no open source
alternative could be found, but which enjoys widespread use in
software engineering research) are available as open source software
for public use, and we have taken care that required libraries and
programming environments are also available under the same
conditions. We have invested substantial effort on documenting not
only the code, but also our development and design choices. The
\(\approx\) 400 commits required for the study are augmented by roughly 130KB of
revision control system notes---about the size of the
actual publication---which not only provides crucial knowledge to
verifiers and replicators of our research, but also facilitates a
simple application of our techniques to other projects. The base data
sets (eMails, issues and revision control data) used for the study are
all available on our supplementary website.


\subsection{Threats to Validity}
\emph{Construct validity.} Since we use social networks constructed
from heterogenous data sources, one threat is that the networks do not
accurately capture reality. Since there is substantial previous
evidence that such networks are authentic in reflecting developer
perception~\cite{Meneelymeasurements,bachelor2014}, this threat is
minor. Another concern regards unification of developer
contributions across multiple data sources to a single identity, and
technical difficulties of parsing real-life data that often fail to
comply with standards. This could lead to networks whose content
differs from the actual content of the data sources. Beside relying
on well-tested analysis code used by multiple research groups over
many years, we subjected our additional code to internal peer review
among the authors, which limits this threat to validity.

Another threat is that the issue and email records are necessarily
incomplete, and that unobserved communication channels would alter the
results. It seems implausible, though, that developers would extensively
use such alternate channels without publicly announcing them (while
email communication may be seen as outdated at the time of writing,
contemporary studies like Ref.~\cite{Ramsauer2019} show that for
projects with eMail based workflows, covert code integration can be
detected by the \emph{absence} of email discussion, which in turn
strongly suggests that all relevant communication is performed via
email for such projects). The accompanying website contains
correlation plots and summary graphs that show results for all measure
combinations resolved by eMail only and eMail+bug tracking data sets
(since the number of eMail messages available for the typical project
relatively exceeds the number of bug tracker entries by usually at
least one order of magnitude, we did not consider bug tracker entries
alone), and little differences can be observed for these two
alternatives.

In the same rein, we rely on the validity of bug and issue tracker
entries in the sense that they need to represent concerns that agree
with the software engineering notion of ``bug'', but whose quality is
known to vary~\cite{Herzig2013}.

While our approach generally meets or exceeds the state of the art for
analyzing socio-technical data, it may be the case that current
approaches are not rich enough to identify strong
relationships between socio-technical congruence and software
quality. For example, representing developer communication
without considering the content of their discussions may obscure
important details. Likewise, the content of a commit may be relevant
to reasoning about socio-technical congruence and its relationship to
software quality.

\emph{External validity.}  We draw our conclusions from analysing a
manual selection of 25 open source projects. Possible consequences of
this manual selection are mitigated by choosing a wide variety of
projects that differ in many dimensions and constitute a diverse
population.  Furthermore, we considered the longest possible
historical data and time-resolved analysis to prevent temporally
biasing our results. The restriction to ``large'' projects with active
histories is less of a threat, because coordination in smaller
projects with only a handful of developers provides few opportunities
for coordination problems. Since open source development tools and
methods are seeing strong application in closed development efforts,
and since a substantial fraction of commercial projects rely on open
source components, the line between open and closed development has
blurred~\cite{Paulson2004,Hunsen:2016,Hofmann:2013}; the threat from choosing to
analyse only open source projects is therefore deemed minor.

\emph{Internal Validity.} Software quality has many aspects, and
focusing on two selected measures does not capture the phenomenon in
its entirety. Since bugs and churn are some of the most impactful and
widely studies observable measures of quality, other measures should
not paint an entirely different quality picture, and it is known that
such indicators are usually highly correlated~\cite{Aranda2009}. To
substantiate this assumption, we have computed a number of complexity
and volume metrics (for instance, average and maximal cyclomatic
complexity, essential complexity, and other measures provided by the
Understand tool) for the subset of projects discussed in
Fig.~\ref{fig:tt_overview}. We find varying \concept and anti-\concept
behaviour without discernible pattern. The absolute magnitudes of
correlation strengths are typically even below those reported for bugs
and churn. While the aforementioned measures are widely used in
practice, there is no universally accepted consensus on which measures
are optimal (or preferable) in which circumstances.  Therefore, we
show details only in graphs in the online supplement, and refrain from
further discussion in the main part of the article. This restriction also
increases internal validity.

Likewise, experiments with high-level global indicators such as the
scalar, global decoupling level~\cite{Mo:2016} are not reported here
for lack of space, but strengthen our conclusion that the described
threats are minor. Calculation results are available in the online
supplement.

There are other formalisations of socio-technical collaboration beyond
the triangle and square motif employed in our approach. However, owing
to the elementary structure of our (anti-) motifs, it is likely that
they would appear as sub-(anti-) motifs of larger (anti-)motifs, which
makes it unlikely that correlations would differ much from what we observe.

\subsection{Sensitivity Analysis}\label{sec:sensitivity}
Our study requires choosing the width of the analysis time window, for
which a certain amount of subjectivity cannot be avoided.  Determining
the relationship between structure (via artefact--ar\-te\-fact
dependencies) and communication obviously requires knowledge of the
structure. For semantic and static dependencies, the structure is
determined from the state of the source code at the beginning of an
analysis interval; inferring the structure via the co-change mechanism
requires analysing \emph{changes to the source} that themselves happen
during the chosen time period. We have chosen a historical window of
one year to gather changes (with the intention of capturing sufficient
data to get an accurate representation of dependencies, and not
capturing too much data to inherit outdated architectural features),
and have verified for a randomly chosen subset of the subject projects
that varying the window size by 3 to 6 months does not alter our
conclusions (additionally, Ref.~\cite{Meneelymeasurements} shows that
enlarging the analysis window beyond three months only
mar\-gi\-nal\-ly changes developer networks extracted from version
control systems).

\section{Discussion}\label{sec:discussion}
Basing statistical analyses on dichotomic, lexicographic decision
rules used to be standard scientific practice over many decades, and
consequently, many of the previous attempts at studying the problem at
hand are based on this paradigm.  We feel it is appropriate to
reiterate that our study is not any more based on such rules, which
follows the latest recommendations of the statistics community.  It
has (once more over decades, but with recently increasing force) come
to the attention of researchers in many fields (see,
\eg,~Ref.~\cite{McShane2019}) that using point null hypotheses that
imply a sharp distinction between effect and non-effect (and also zero
systematic error) are highly problematic~\cite{Berkson1938, Bakan1966,
  Tuckey1990, Gelman2015}. They describe overall implausible
situations. Since a larger number of smaller, noisier studies is
additionally bound to detect statistically significant results with
higher probability than a single large-scale study~\cite{simmons2011},
we have deliberately opted for the latter, accepting the many problems
that this raises in presenting, discussing, and publishing the
results. However, we find the approach leads to a very clear
picture.

\subsection{Impact on Structure, Organisation and Priorities in
  Large-Scale Software Development}
To reiterate our major empirical finding: We have, using a multitude of
carefully constructed models from different statistical schools of
thought, observed only negligible relationships between \concept and
our measures of software quality: bugs and churn. And, perhaps more
importantly: The \emph{direction} of any possible relations varies
randomly over time, such that the very same socio-technical scenario
that leads to good outcomes at one point in time can lead to bad
results in the next. What conclusion should we draw from these results
which, it must be emphasized, contradicts many assumptions about socio-technical congruence
over the past five decades?  

Our interpretation is striking and unambiguous: \concept, at least as it is manifested in our motifs and anti-motifs, does not
matter with respect to the number of bugs in the code and the
associated densities, nor with respect to code churn; or at least does
not matter much, as
evidenced by the 25 projects that we studied.

If there is a measurable relation between \concept and properties of
software artefacts, the relation must be manifested at a level of
abstraction that is higher than relations between pairs of software
files.  The key practical consequence, contrary to commonly made
assumptions, is that software architects and project managers of
large, complex software systems should not spend too much time on
optimizing the communication structures of such projects with bug
density and churn in mind.  While these social aspects of projects are
obviously not unimportant, they just as obviously do not have a strong
impact on critical, measurable project outcomes.  Given that every
project is budget and schedule constrained, project leaders should
therefore be spending more of their finite resources on improving
other aspects, for example: knowledge of tools and languages,
completeness and automation of testing, coding practices, automation,
and so forth.

\subsection{Relation to Landmark Investigations of Socio-Technical Congruence}
The notion of socio-technical congruence was shaped by Cataldo et
al.~in Ref.~\cite{Cataldo2008} and subsequently refined by the same
authors~\cite{CataldoH13} where they elaborate socio-technical
congruence in formal terms and empirically investigate its impact on
product quality. Similar work~\cite{rel9} led to the many
network-based declinations of socio-technical congruence, for
instance, its evolution into a network-based software community
awareness~\cite{awarenets}, in which de Souza et al.\ discuss it as the
basis for awareness maintenance mechanisms. Since the seminal works of
Cataldo, Herbsleb and colleagues discuss questions that are similar to
ours (the impact of various notions of socio-technical congruence on
key software quality indicators) and use related techniques
(multivariate linear and logistic regression), but arrive at different
conclusions, it seems pertinent to comment in more detail on the
relation between their findings and the conclusions of this investigation.

Firstly, the types of socio-technical congruence used in the different
studies are related, but not identical: Cataldo~\etal define
socio-technical congruence via technical dependencies based on either
static coupling (data/function/method references across files) or
alternatively, in our terminology, co-changes~\cite{hg99, CataldoH13,
  Cataldo2008}. Secondly, the aforementioned studies consider file
buggyness (has a file been modified in the course of resolving a field
defect?) and resolution time for change requests as main quality
indicators respectively covariates of interest; the conceptual
relation to bug density is obvious, but the measures are not the
same. Regarding statistical power, the base data for the cited studies
cumulatively encompass 2 projects, 8 development years, and 154
developers, which is, at least, an order of magnitude less than our
dataset in every aspect.

Given such pronounced conceptual similarities, how can it be that
Cataldo et al.\ arrive at conclusions along the lines of \emph{``all
  (...)  types of dependencies are relevant and their impact is
  complementary, showing their independent and important role in the
  development process''~\cite{Cataldo2008}}, while we find that
socio-technical congruence is of less importance than previously
believed? Fortunately, both points of view can be reconciled if the
focus is not put on statistical \emph{significance}, but on
\emph{effect size} and \emph{relevance}, which is also what matters
in the end of the day in practical software development.

The regression models constructed by Cataldo~\etal~\cite{Cataldo2008} relate
resolution time with many regressors including various forms of
\emph{congruence}. The achieved \(R^{2}\) values lie around 0.75,
which is similar to our models, and \(p\) values for the
congruence-related covariables are between 1\% and 10\%, which
is---even by proponents of the use of \(p\) values---usually taken as
not quite excessively significant. Depending on the time interval we
consider in our models, it can well happen that we arrive at similar
\(p\) values, although our interpretation is different, following the
statistical approach that we have discussed at length in this article.
Of course, \(p\) values are uniformly distributed under the null
hypothesis, and values in this range are not too unlikely to arise for
large enough samples under \(H_{0}\).

While the exact interpretation of \(p\) values could be seen as
statistical fine print by some when the widely deployed
dichotomisation into significant and non-significant contributions is
employed, a perhaps even more important consideration concerns the
\emph{relative magnitude} of regression coefficients. Coefficients
that deal with congruence are substantially smaller than those for
other influence factors, and differ in some cases even by orders of
magnitude. As a concrete example, and again referring to the study of
Cataldo~\etal~\cite{Cataldo2008}, this is most pronounced for change
request priority (coefficient -0.4), but other covariables (for
instance, change size with a coefficient magnitude of 0.31) also fall
into this class.  Contrast this to a coefficient of -0.05 for
structural congruence: If, for instance, the priority of a change
request is doubled, the relative contribution to resolution time is
-0.8, and the request will be processed substantially quicker (as it
should be). If the amount of structural congruence is doubled, the
relative contribution is only -0.1, which is drastically
smaller. Considering that changing the priority of a request can
usually be performed with the literal ``mouse click'', a substantial
effort in terms of work and team organisation can be expected to be
required to make any change to the socio-technical structure of a
project. Consequently, we find that many realistically achievable
effects of socio-technical congruence are more likely to be on the
level of perturbative variations of the major influence factors. It
goes without saying that an identical conclusion can be drawn from our
results, too.

Consequently, we argue that there is no unresolvable disagreement
between our study and previously achieved seminal results in terms of
statistical inference, but predominantly in terms of
\emph{interpretation} of these results.  To arrive at a strong
conclusion, it is of course necessary to perform a careful
mixed-methods analysis (to not introduce bias by a particular method) on
a sufficiently large sample size (to achieve generality), including a
careful sensitivity analysis (to combine different viewpoints), as we
have done in our study.

\subsection{Relation to other Socio-Technical Hypotheses}
The relationship between social and technical aspects of software
development has been discussed and investigated for about half a
century (e.g., consider for example, Conway~\cite{Conway:1968} as well as others after him~\cite{herbsleb1999beyondconwayslaw,BirdPDFD08,Tamburri19}). Conway's Law, for instance, is the
earliest in a family of socio-technical hypotheses that relate
software structure to the organizational structures producing it. It
first appeared in 1968~\cite{Conway:1968} as an empirical observation
that attempted to relate the structure (or, in modern terms,
\emph{software architecture}) of a system with the structure of the
\emph{organization} that creates and maintains it~\cite{ossslr,dsn}:
``\emph{Organizations which design systems [\ldots] are constrained to
  produce designs which are copies of the communication structures of
  these organizations.}''

From a simple textual analysis, Conway's Law postulates: (1) a
relationship between software system structure and its
communication/organisational structure~\cite{ossslr}; (2) that
organisations are \emph{constrained} by some invisible force to
produce designs that mirror the organisational structure. It seems
immediately evident that this notion can be easily mapped to our
definition of \concept, or is, at least, closely related to it.

What the law, however, does not explain is what might occur if this
constraint is violated (a law without non\hyp{}trivial consequences is
of limited practical and theoretical value).  This shortcoming
severely hampers the degree to which the law can be used as an
empirical device to guide the organization of large\hyp{}scale
software engineering projects and their architectures: The function of
a law in scientific considerations is to describe, in the best case
quantified and mathematically, a generalised observation on how
certain things relate to each
other~\cite{Winther2016,Andersen2016}, given a particular
theoretical framework that explains one or many laws for the smallest
possible number of requires observable quantities. In the process of
advancing scientific progress, it is natural that measurements can
appear at some point in time that violate a given law, which then
prompts the development of either extended theories, or to reformulate
laws. Of course, the impact of any hypothesis or law without
observable consequences that differ depending on whether it is
fulfilled or not is limited, and our study suggests that this scenario
seems to apply for \cl.

Regardless of this limitation, \cl is intriguing and potentially
far-reaching in impact and meaning, and therefore it has been the
subject of numerous studies that address aspects such as its
implication for distributed software de\-ve\-lop\-ment~\cite{BSF+13},
software tasking~\cite{KwanCD12}, splitting complex software
organizations~\cite{hg99}, or how designs mirror a project's
communication structures~\cite{colfer2010mirroring}, to name a
few. However, we feel that our work shows that verifying or falsifying
the law is actually not a pressing issue--- for the simple reason
that, following our results, it does not matter much whether it holds
or not in terms of practical software engineering and software quality
consequences.

\subsection{Limitations}
There are clearly some limitations to our definition of
socio-technical motifs, their relation to software quality, and the
interpretation of connections and non-connections discussed in this
study. The projects that we chose are all fairly large, with dozens to
hundreds of contributor, and thousands of issues and commits. And
these projects all have ``average'' levels of overall complexity (as
can, \eg, be derived from global measures such as the decoupling
level~\cite{Mo:2016}, which we computed for all subject projects and
found the results to fall in the ``average'' range \([0.25, 0.7]\) of
the unit interval), which is to say that their architectures are
neither ideal nor are they toxically bad. One might expect that, in
projects with highly coupled architectures, communication would be
absolutely essential to manage the complexity.  One of the major
purposes of decoupling, which we typically achieve through abstraction
and the application of patterns, is to allow for the independent
activities of developers.

We conclude that the next step of our evidence-based endeavour into
the nature of \concept will need to involve projects with very high
(or low) levels of coupling, at the tails of the distribution, where
we might observe a stronger influence of \measure. Additionally, we
will need to investigate if other quality measures and indicators
(see, \eg,~\cite{Hall:2012}) exhibit a stronger connection with
\concept than the measures used in this work.



\section{Conclusion}\label{sec:conclusion}
In this article we have presented a method to empirically investigate
the connection between the presence of socio-technical structural
observables with software quality as determined by robust, practical
measures based on elementary network motifs that formalise one
particular, yet fundamental notion of socio-technical congruence. A
large-scale longitudinal study on a diverse set of software projects
has shown that these motifs occur strongly non-randomly, and that
their occurrence varies as the projects evolve.  We have defined a
quantitative and interpretable notion of socio-technical motif
congruence, and have shown that it is, in no substantial way, related
to measurable project quality outcomes---soft\-ware bugs, bug density
and churn---in any temporal scenario.

A key lesson learned is that socio-technical congruence has less substantial consequences than was
previously believed, and hence might not deserve the great attention
that it has received. Our argument can be extended to hypotheses like
\cl that have received great attention during the last five decades,
but often fail to result in appreciable, measurable and quantifiable
consequences.


\section*{Acknowledgments}
Rick Kazman was supported in part by the National Science Foundation
of the US under grant CCF-1514561. Sven Apel has been supported by the
German Research Foundation (AP 206/14-1). Wolfgang Mauerer
acknowledges mobility support from grant BayIntAn-OTHR-2-16-144, and
support from Siemens AG, Corporate Technology. This work was supported
by the iDev40 project. The iDev40 project has received funding from
the ECSEL Joint Undertaking (JU) under grant no.~783163. The JU
receives support from the European Union’s Horizon 2020 research and
innovation programme. It is co- funded by the consortium members,
grants from Austria, Germany, Belgium, Italy, Spain and Romania.
The authors thank Drexel University (Qiong Feng and Yuanfang Cai) for
providing a Jira issue crawler.\flushcolsend

\begin{IEEEbiography}[{\includegraphics[width=1in,height=1.25in,clip,keepaspectratio]{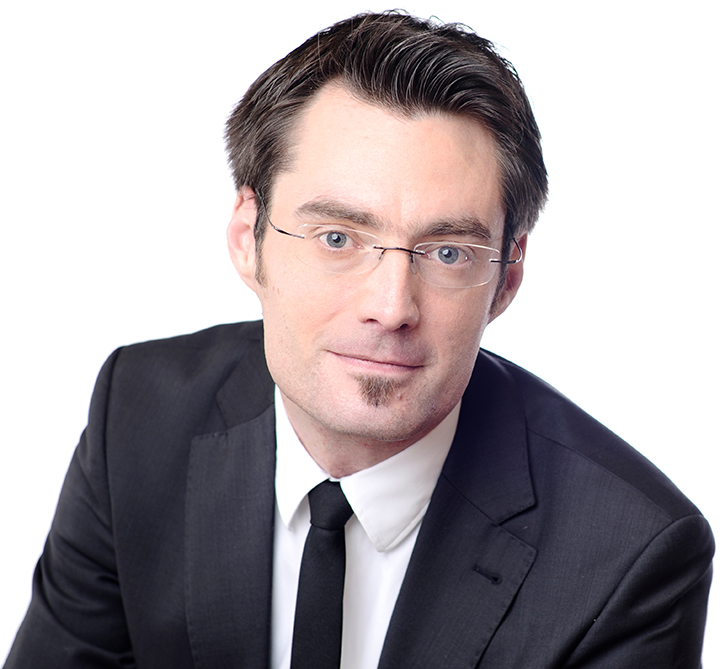}}]{Wolfgang
    Mauerer} is a Professor of Theoretical Computer Science at the
  Technical University of Applied Sciences Regensburg, and a Senior
  Research Scientist at Siemens AG, Corporate Research, Munich. His
  interests focus on quantitative and empirical software engineering,
  low-level systems engineering, and quantum computing. He
  received his PhD from the Max Planck Institute for the Science of
  Light. Contact him at \url{wolfgang.mauerer@othr.de}.
\end{IEEEbiography}

\begin{IEEEbiography}[{\includegraphics[width=1in,height=1.25in,clip,keepaspectratio]{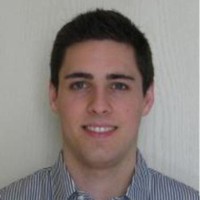}}]{Mitchell
  Joblin} is a research scientist at Siemens AG, Corporate
Research, Munich and a postdoctoral researcher at Saarland
  University. His research interests include empirical software engineering, software analytics, network analysis, and machine learning on heterogeneous information networks. He received his PhD in Computer Science from the University of Passau.
Contact him at
\url{mitchell.joblin@siemens.com}.
\end{IEEEbiography}

\begin{IEEEbiography}[{\includegraphics[width=1in,height=1.25in,clip,keepaspectratio]{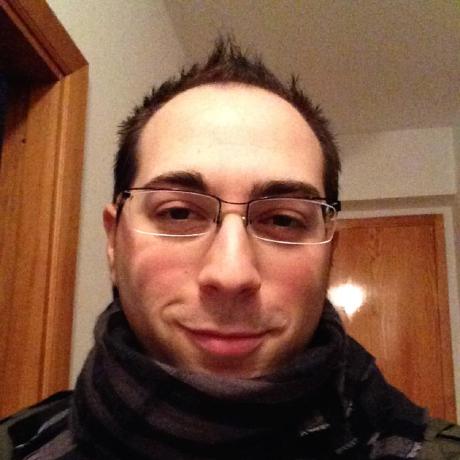}}]{Damian
    A. Tamburri} is an Associate Professor at TU/e -- JADS. His research
  interests include social software engineering, advanced software
  architecture styles, and advanced software-architecting
  methods. He's on the IEEE Software editorial board and is secretary
  of the International Federation for Information Processing Working
  Group on Service-Oriented Computing. Contact him at
  \url{dtamburri@acm.org}.
\end{IEEEbiography}

\begin{IEEEbiography}[{\includegraphics[width=1in,height=1.25in,clip,keepaspectratio]{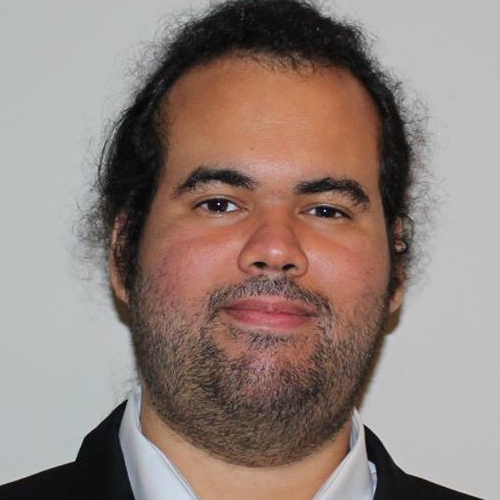}}]{Carlos
    Paradis} is a doctorate student and Research Assistant at the
  University of Hawaii. His primary research interests are software
  vulnerabilities, safety, text mining, and text visualization
  methods. Paradis has worked on several data-driven projects,
  including software engineering, aerospace, renewables, thermal
  comfort, and healthcare. He is also a member of IEEE and ACM’s honor
  societies. Contact him at \url{cvas@hawaii.edu}.
\end{IEEEbiography}

\begin{IEEEbiography}[{\includegraphics[width=1in,height=1.25in,clip,keepaspectratio]{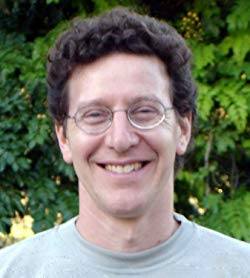}}]{Rick
    Kazman} is a Professor of Information Technology Management at the
  University of Hawaii and a principal researcher at Carnegie Mellon
  University's Software Engineering Institute. His research interests
  include software architecture design and analysis tools, software
  visualization, and software engineering economics. Kazman received a
  PhD in computer science from Carnegie Mellon University. Contact him
  at \url{kazman@hawaii.edu}.
\end{IEEEbiography}

\begin{IEEEbiography}[{\includegraphics[width=1in,height=1.25in,clip,keepaspectratio]{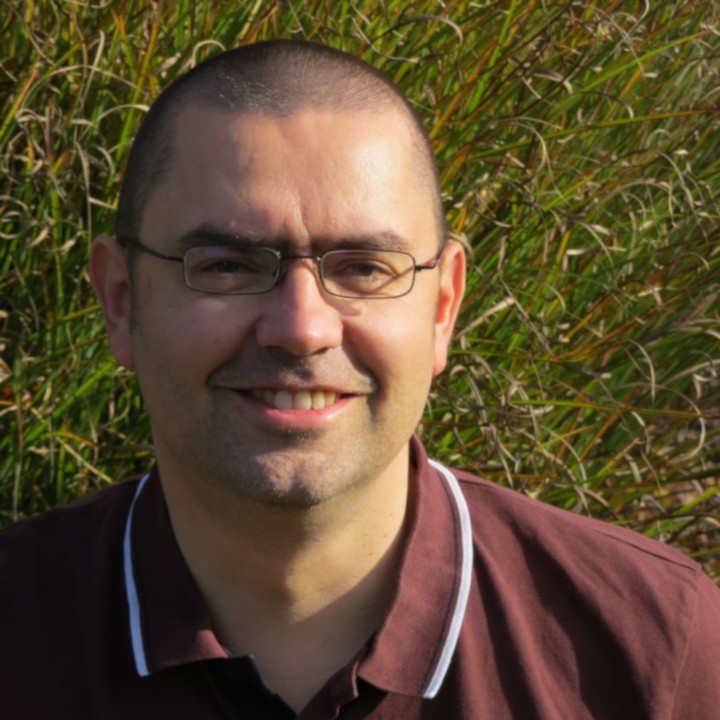}}]{Sven
    Apel} holds the Chair of Software Engineering at Saarland
  University \& Saarland Informatics Campus. His research interests
  include software product lines, software analysis, optimization, and
  evolution, as well as empirical methods and the human factor in
  software engineering. He received a PhD in Computer Science from the
  University of Magdeburg. Sven Apel is an ACM Distinguished
  Member. Contact him at \url{apel@cs.uni-saarland.de}.
\end{IEEEbiography}

\vspace*{\fill}\newpage

\section*{Appendix}
\subsection*{Validity of Socio-Technical Base Data}\label{sec:validity_base_data}
\begin{figure*}[htbp]
  \incfig{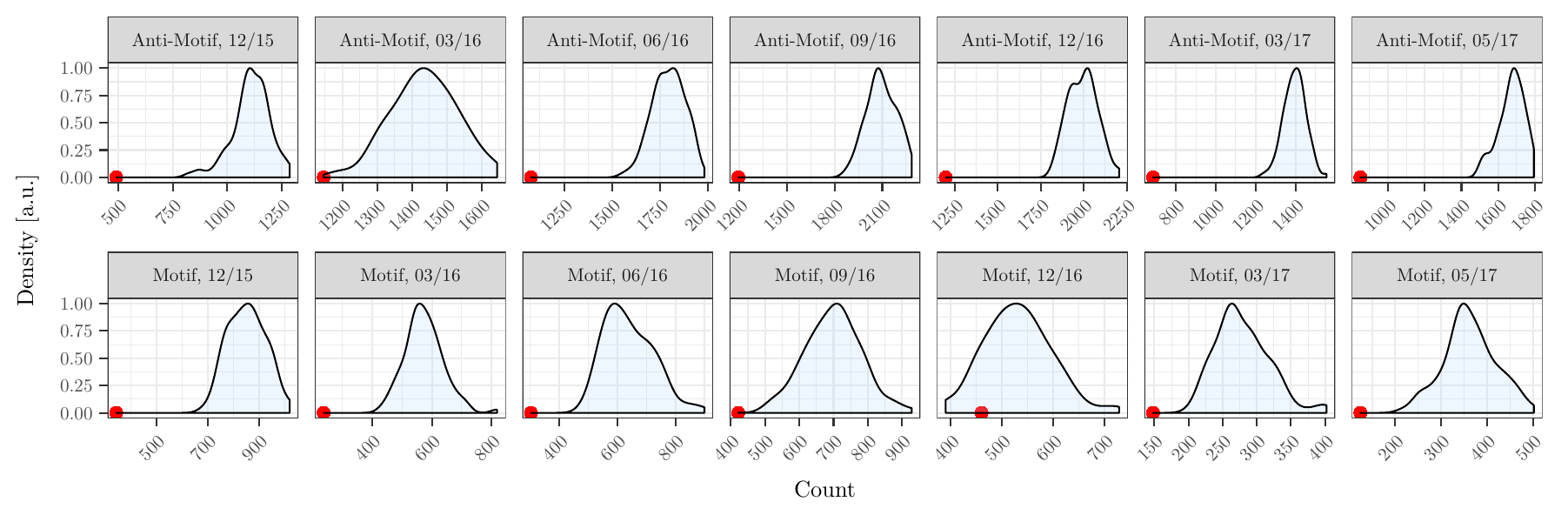}\vspace*{-1em}
  \caption{Empirical motif count (red dot) and count distribution 
    given by the rewiring process illustrated for a subset of the 
    complete analysis period for project \HBase. The calculation 
    ensures that the networks our considerations are based on 
    represent meaningful information in the data.}\label{fig:null_model_ts}
\end{figure*}

We have discussed in Section~\ref{sec:config_model} how the
base data for our analysis are gathered, and how they are processed
into a bi-model graph that represents the socio-technical network.
We elaborated on page~\pageref{sec:config_model} how we ensured that
the generated graphs are congruent with the real-world structure of developers
and artefacts. Figure~\ref{fig:null_model_ts} shows a time-resolved
graph that visualises the results of the configuration model
hypothesis testing procedure resolved by three month time windows for project 
\HBase for the square motif and square anti-motif. This figure shows 
probability density functions for the possible different 
(anti-) motif counts \(p_{m,t}(n)\) of a network generated randomly 
with the same degree sequence as the real world network. The red dot 
indicates the project's observed, empirical motif count \(c_{m,t}\). 

It is visually imminent that the observed count is extremely unlikely
to stem from any of the simulated distributions, albeit this can also
be statistically ascertained by using a t-test that is significant at
(for all practical purposes) arbitrarily small levels for almost all
temporal ranges. In a small fraction of all temporal ranges (for
instance the interval ending in 12/16), the real data are partly
compatible with a random network structure. Such outliers are not
unexpected for noisy real-world data. We did not try to investigate
exact causes for these situations (that could, for instance, be caused
by erratic and uncoordinated intermittent phases in a project's
lifecycle), but we have instead made sure that subsequent analysis stages
are sufficiently robust to also deal with such base inputs.

\subsection*{Model Assumptions and Conditions}\label{sec:model_preconditions}
In Section~\ref{sec:multivariate_regression} on
page~\pageref{sec:multivariate_regression}, we have introduced linear
regression models to analyse the dependence of software quality
indicators, bug density and churn, on the socio-technical motif
structure in the presence of other influence factors. Valid linear
models need to satisfy various assumptions and conditions, and we
analyse next how well this holds for our models and data.

\begin{figure}[htbp]
  \vspace*{-3em}\incfig{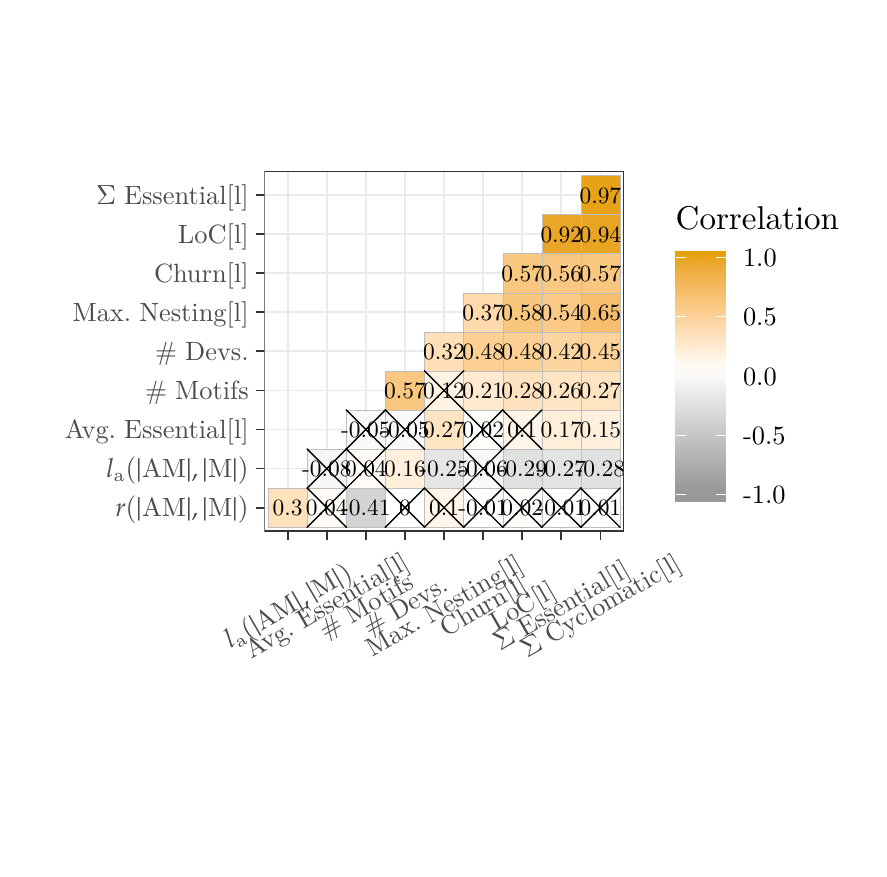}\vspace*{-8em}
  \label{fig:correlations_overview}\caption{Correlogram to detect
    multi-collinearity between candidate regression model covariates.
    One randomly chosen temporal range (28) of project \HBase is used
    to illustrate the interdependence of possible predictors, but
    other projects and subsets exhibit a similar structure and
    correlation values. Colour of the table fields entries indicates
    sign and magnitude of the measured correlation, crossed out fields
    mark insignificant values (at the usual 5\% significance level).}
\end{figure}

Figure~\ref{fig:correlations_overview} shows the correlation structure
of the measured variables that are available for our study. Apart from
quantities like bug density, churn, motif count and \measure that form
the core of our study, we also include traditional complexity
indicators like (sum/average) cyclomatic/essential
complexity~\cite{McCabe1976} and maximal nesting~\cite{Harrison1981} to allow for
assessing the relative influence of novel metrics in comparison to
more traditional ones. The latter possess know weaknesses that are 
also reflected in the diagram---sum\footnote{\emph{Sum} refers to
  the sum of all essential complexity values for every function in a
  given artefact; average complexities are computed by normalising
  said value by the number of functions in an artefact.}  of essential
complexity, sum cyclomatic and file size in terms of lines of code are
highly correlated. To avoid collinearity in the model, yet include
traditional software engineer that can serve as a ``baseline'' and
comparison measure on the usefulness of \measure, we exclude the sums
of essential and cyclomatic complexity, and keep the admissible (in
the sense of aptitude for linear models) measures \emph{maximal
  nesting} and \emph{average cyclomatic} complexity.

\begin{figure}[htbp]
  \incfig{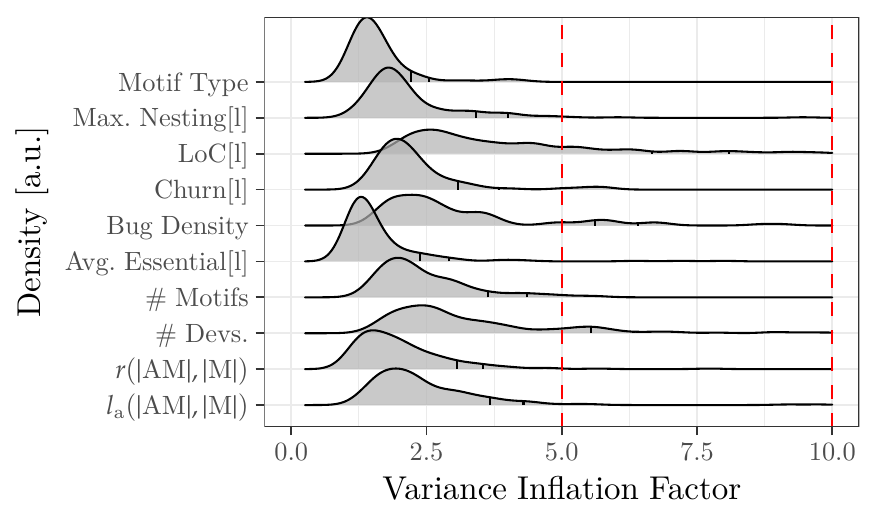}\vspace*{-1em}
  \caption{Variance inflation factor distributions for all
    covariates contributing to bug density and churn models, and
    encompassing all temporal ranges for \Spark, \HBase,
    \TrafficServer, \Groovy, and \Cassandra. Vertical solid lines
    embedded in the densities mark the 90\% and 95\% quantiles; dashed
    lines from top to bottom indicate two commonly used upper bounds
    for unproblematic values of VIF.}\label{fig:vif_global}
\end{figure}

The correlation values shown in Figure~\ref{fig:correlations_overview}
reflect the situation for a single temporal range of a single subject
project. To establish an appropriate set of regressors for the full
dataset (it is not only impractical to specify different models
depending in project and time range, but would also limit
comparability of the resulting insights), we resort to computing the
distribution of the \emph{variance influence factor} (VIF) for all
projects that are subjected to regression analysis. The VIF is
computed for each regressor per project and analysis time range, and
results (for this combination) in a single scalar value that
quantifies the amount of collinearity of the regressor with the other
covariables.  Figure~\ref{fig:vif_global} presents the results. While
there is no accepted strict threshold for VIF above which a variable
is considered ``too'' collinear, strictly applying rules of thumb is
problematic~\cite{OBrien2007}, but values below five or even ten are
usually regarded as inconsequential~\cite{Hair1995}.  The regressors
selected in the above consideration do in the vast majority of all
cases not exceed either threshold, and can thus be used to specify
one single, unified project- and time independent regression model.


\subsection*{Details of Data Collection}
We have outlined the general mathematical structures and components
used to represent developer, artefact and developer-artefact-networks
in Section~\ref{sec:network_construction}. Since they are based on
data obtained from real-world systems, constructing them is a complex
and intricate, yet mostly technical issue. We provide more details on
the required steps in this appendix, but want to explicitly point out
that the source code of our analysis pipeline is the definitive
reference for all data processing and transformation operations required to
replicate our findings.

Information concerning development artefacts is extracted from data in
version control systems. Our pipeline supports the analysis of git
repositories, which does not limit generality of our approach because
nearly every other VCS can be converted into a git repository without
loss of information. Git supports highly non-linear histories based on
multi-branch development, and the first step is to linearise this
history into one time-ordered sequence of events based on when a
commit was authored. We rely on capabilities provided by git itself
for this purpose, like in most previous work.

The linearised commit sequence is then traversed for each time
window, and snapshots of the source tree for each commit are used as
basis for constructing the various artefact-artefact dependencies:

\begin{enumerate}
\item \emph{Static dependencies} are computed by running the
  ``Understand'' tool from SciTools (build 838), exported into CSV format
  by \texttt{und export -dependencies file csv} (the exact call
  sequence is given in file \texttt{codeface/R/gen\_dsm.r}), and then
  converted into a dependency structure matrix (DSM) by our analysis
  pipeline. The Understand tool computes all dependencies that relate a
  (sub-)artefact in one file with a (sub-)artefact in another. For
  instance, if a function in file \(A\) calls another function in file
  \(B\), a dependency between \(A\) and \(B\) arises. Other
  dependencies taken into account include textual imports
  (for instance, header files), and inheritance relationships for
  object-oriented languages; a full list is given on the product
  website.  From the set of languages employed in the subject
  projects, Understand is capable of parsing C, C++, C\#, Python, and
  Java. Projects Ambari and CouchDB, which are written in
  JavaScript and Erlang, respectively, are excluded from the
  DSM-based analysis. While Understand creates a weighted DSM, our
  analysis does not consider the ``strength'' of a dependency, only
  its existence.
\item \emph{Evolutionary dependencies}, also called co-changes, are
  directly inferred from our analysis pipeline (the concrete
  implementation is given in function \texttt{query.dependency} in
  \texttt{codeface/R/dependency\_analysis.r}). By iterating over the
  linearised list of commits, we infer which file artefacts are
  touched for each commit. Using this information, and given a
  specific file \(f\), we can then infer the set of files
  \(\{f'_{1}, f'_{2}, \ldots, f'_{N}\}\) that changed jointly with
  \(f\) in any of the previous commits. The type of change (addition
  or deletion) is not taken into account.
\item \emph{Semantic dependencies} are the most involved coupling
  mechanism to compute. For a given snapshot of the source code under
  consideration, our pipeline first extracts the implementation for
  each function in the system, (code and comments); the boundaries of
  each function within a file are obtained using the Doxygen tool.
  Multiple established text mining (TM) steps are then applied to each
  function (\emph{document} in TM terminology):
  \begin{itemize}
  \item \emph{Stemming} reduces word diversity by removing suffices
    (\eg, ``ing'', ``ly'', ``er''), thus bringing words to their root
    form, and eliminates words that contain little information.
  \item A \emph{term-document matrix} (TDM) of size \(M\times N\),
    where \(M\) is the number of keywords and \(N\) the number of
    documents, is created from the stemmed data. The entry at
    position \((i,j)\) is non-zero when document \(d_{j}\) contains
    term \(t_{i}\).  To increase the influence of terms that describe
    distinct concepts and decrease the influence of the remaining
    terms, we apply an established standard weighting scheme, the
    \emph{frequency-inverse document frequency} (Ref.~\cite{Joblin:2017} gives a detailed
    rationale for this choice).
  \item \emph{Latent semantic indexing}, a matrix decomposition
    technique that relies
    on the singular value decomposition, is used to project the
    high-dimensional space of employed terms into a much
    lower-dimensional subspace, with the added benefit of resolving
    synonymic and polysemic relationships. The technique is a standard
    approach described in detail in~\cite{Baeza-Yates:2000}, and has
    been shown to be applicable and useful for the problem at hand by
    Bavota et al.~\cite{Bavota:2013}.
  \item Semantic coupling between documents is obtained by computing
    the cosine similarity between all document vectors projected onto
    the lower dimensional subspace attained from applying latent
    semantic indexing. Two documents are considered to be coupled if
    the coupling value exceeds a certain threshold, for which we use a
    value suggested by Joblin et al.~\cite{Joblin:2016} after
    extensive experimentation.
  \end{itemize}

  Finally, the results obtained on a per-function basis are then
  aggregated to a per-file basis to ensure alignment with the
  analysis granularity used for the other coupling mechanisms.
  \end{enumerate}

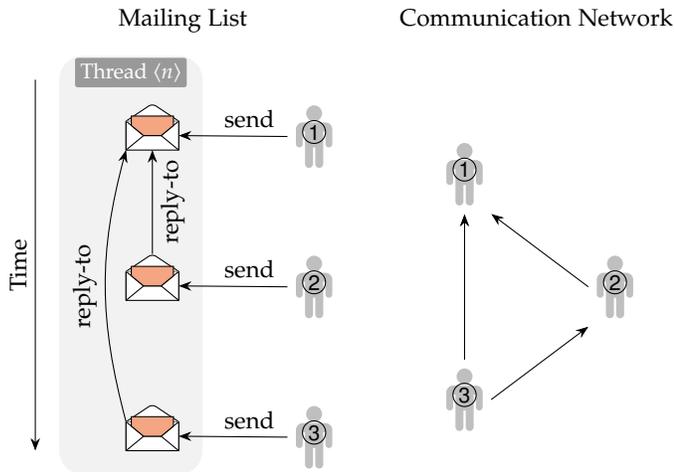
\begin{figure}[htbp]
  \begin{center}
      \def\emwidth{0.7cm}
  \def\pshift{-0.35cm}
  \def\cshift{-0.1em}
  \def\pdist{-0.35cm}

  \begin{small}\begin{tikzpicture}

    \envelope{1,4}{\emwidth}{em1}
    \envelope{1,2}{\emwidth}{em2}
    \envelope{1,0}{\emwidth}{em3}

    \newcommand{\pernum}[1]{\circled{\footnotesize\textsf{#1}}}
    \node[anchor=south,yshift=\pshift] (p1) at (3.5,4) {\manschgerl}; 
    \node[anchor=south,yshift=\cshift] (c1) at (3.5,4) {\pernum{1}}; 
    
    \node[anchor=south,yshift=\pshift] (p2) at (3.5,2) {\manschgerl}; 
    \node[anchor=south,yshift=\cshift] (c2) at (3.5,2) {\pernum{2}}; 
    
    \node[anchor=south,yshift=\pshift] (p3) at (3.5,0) {\manschgerl}; 
    \node[anchor=south,yshift=\cshift] (c3) at (3.5,0) {\pernum{3}};
    
    \path[-Stealth] (em2/top) edge node [below,rotate=90] { reply-to } ($(em1/bl)!0.5!(em1/br)$)
    (em3/tl) edge [bend left=15] node [above,rotate=90] { reply-to } (em1/bl)
    ;

    \path[-Stealth] (p1.west) edge node [above,pos=0.35] { send } (em1/cr);
    \path[-Stealth] (p2.west) edge node [above,pos=0.35] { send } (em2/cr);
    \path[-Stealth] (p3.west) edge node [above,pos=0.35] { send } (em3/cr);

    \coordinate(boxtl) at  ($(em3/bl |- em1/top) + (-1.8em,1em)$);
    \coordinate(boxbr) at (em1/tr |- em3/br);
    \begin{pgfonlayer}{background}
      \filldraw [line width=2em,join=round,black!5]
      (boxtl) rectangle (boxbr);
    \end{pgfonlayer}
    \node[piclabel] [above=of em1/top,anchor=center,yshift=-1.85em,xshift=-0.9em]
    { \glab{Thread \(\langle n\rangle\)}};

    \path[-Stealth] ($(boxtl) + (-2em,0)$) edge node [above,rotate=90] { Time }
    ($(boxbr -| boxtl) + (-2em,0)$); 

    \node[align=center,anchor=north] at (1.8,6) { Mailing List };
    \node[align=center,anchor=north] at (6.5,6) { Communication Network };


    \node[anchor=south,yshift=\pshift] (p1) at (5.5,3.5) {\manschgerl}; 
    \node[anchor=south,yshift=\cshift] (c1) at (5.5,3.5) {\pernum{1}}; 
    
    \node[anchor=south,yshift=\pshift] (p2) at (7.5,2) {\manschgerl}; 
    \node[anchor=south,yshift=\cshift] (c2) at (7.5,2) {\pernum{2}}; 
    
    \node[anchor=south,yshift=\pshift] (p3) at (5.5,0.5) {\manschgerl}; 
    \node[anchor=south,yshift=\cshift] (c3) at (5.5,0.5) {\pernum{3}};

    \path[-Stealth] (p3.north) edge node {} (p1.south)
                    (p3.east) edge node {} (p2.south west)
                    (p2.west) edge node {} (p1.south east)
                    ;
  \end{tikzpicture}
  \end{small}
  \end{center}\vspace*{-2em}
  \caption{Constructing developer networks from eMail communication.}
  \label{fig:email_network}
\end{figure}

Communication relationships can be inferred from mailing list and
issue tracker communication. Communication networks are inferred in
the same way for both, since all eMails and issues are associated with
(a) a unique identifier, (b) a person sending/submitting the
mail/issue, (c) a date when the message/issue was sent. Also, the
resulting initiation/response structures can be represented by trees
in both cases. The construction methodology follows standard practises
of the field (see, \eg, Refs.~\cite{BirdPDFD08,Panichella:2014}), and is illustrated in
Fig.~\ref{fig:email_network}: A conversation from a mailing list archive is
shown for three individuals that compose three eMails in a single
thread. Links between individuals and the eMails they have authored
are available in addition to links between eMails that express
``Reply-to'' relationships. The corresponding developer communication
network stemming from activities on the mailing list is given on
the right-hand side of the figure.


Raw data for eMails are obtained from public archives offered by the subject
projects, and issues are downloaded from Jira bug trackers using the
Titan tool (we have filtered for issue type ``Bug'' for the latter
source). All raw data sets are provided on the accompanying website.

While we have tried to provide the most important details of our
technical approach, there is a large number of remaining details that
we cannot comprehensively discuss here. Any technical deviations from
the above that might remain are implicitly documented (and preserved
in a replicable way) by the pipeline source code available on the
aforementioned website.

\subsection*{Model Correctness}\label{sec:model_correctness}\balance
Multivariate linear models (as we consider in
Section~\ref{sec:multivariate_regression} on
page~\pageref{sec:multivariate_regression}) are numerically stable,
and can be very well interpreted when certain model assumptions are
satisfied. Drawing \emph{correct} conclusions is, in general, also only
possible for correctly specified models.  We have taken care to
perform the required checks~\cite{Fahrmeir2013} for the subset of
projects discussed in this and the following section, and need to
point out two problematic aspects inherent in the base data set that
is visible in Figure~\ref{fig:bug_density_residuals} for a randomly
chosen temporal analysis interval. Deviations from the expected
distributions could of course also be detected by formal tests;
following~\cite{Anscombe1973}, we do prefer graphical illustrations
since they provide more fine-grained insights into the nature of the
problems than binary significance tests, and also reduce sensitivity
considerations.

Firstly, there are clear deviations from the
normality assumption \(\vec{\epsilon}\sim \mathcal{N}(0, \sigma^{2})\) for the
residual distributions. Second, we see a substantial amount of
temporal correlation between residuals. While the first issue has no
influence on parameter estimation itself, any derived \(p\) values
will be too low, and confidence intervals too wide, which essentially
implies that the importance of \concept will be even
\emph{overestimated} by our model.  Since we find the influence to be
limited anyway, the influence of mis-specifications on our conclusions
is limited.  Let us also remark that random interval selection was
performed by incrementally seeding a random number generator with a
fixed value that is incremented by one for each time interval
selected. We use this procedure to avoid any (even inadvertent) bias
towards ``welcome'' results in selecting displayed subsets.

\begin{figure*}[htbp]
  \incfig{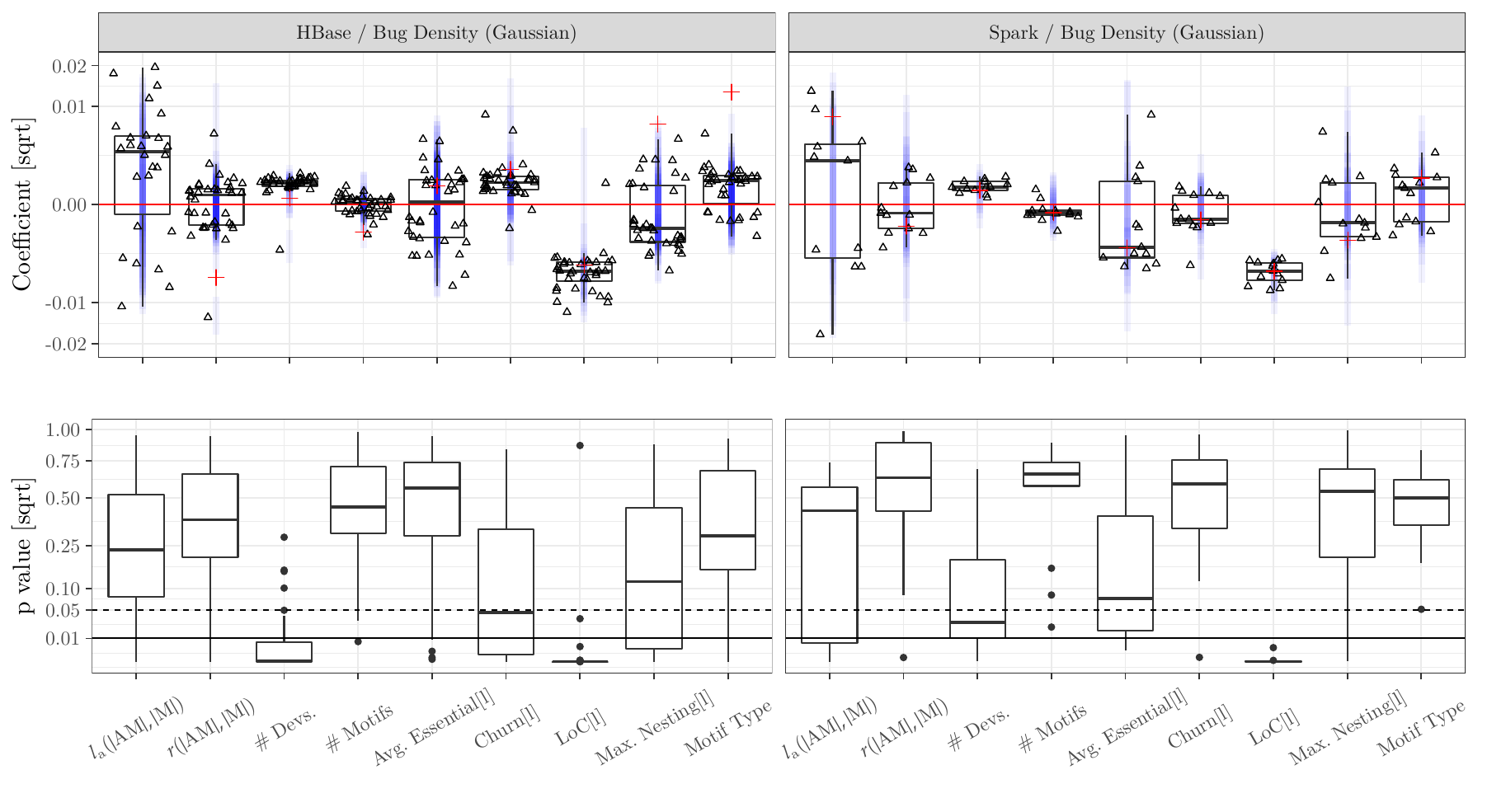}\vspace*{-2em}
  \caption{Results of a time-resolved multivariate linear regression  
    for bug density (the figure is restricted to projects \Spark and  
    \HBase, but results for other projects -- as shown in the online  
    supplement -- are structurally very similar). The interpretation of  
    the graph is identical to Figure~\ref{fig:logistic_regression}:  
    The box plot for every covariate summarises the contributions for  
    all temporal analysis ranges. Red crosses represent the  
    coefficient values resulting from the mixed effects linear  
    regression model~\eref{eq:bug_density_lmm}, and vertical blue  
    lines of varying intensity indicate the 95\% confidence  
    intervals.}\label{fig:bug_density_regression}
\end{figure*}

The problem of correlated residuals can be pinpointed to a particular
structure of the data, as indicated by the distinction between data
points contributed by artefacts with zero bug density
(yellow/triangles), and non-zero such entries (black/dots): The large
majority of artefacts is not associated with any bug. This problem
(or, rather: structural observation) is inherent in the data, and is
directly linked to the way how projects collect bugs, and maintain the
information, which we cannot influence. Consequently, we need to
accept that bug tracker entries are only a proxy for buggyness.  We
have performed a second iteration of all analyses presented in this
article with included zero bug densities, leading to essentially
identical statements as in the cleaned case, but, of course, suffering
from the consequences of violating the model assumptions. Results are
shown in the online supplement. \balance

\bibliographystyle{IEEEtran}
\clearpage\bibliography{main}
\end{document}